\pdfoutput=1
\documentclass[camera,pageno]{jpaper}

\usepackage{multirow}
\usepackage{mathtools}
\usepackage{xspace}
\usepackage{textcomp}
\usepackage{algorithm}
\usepackage{algorithmic}
\usepackage{subfig}
\usepackage{xfrac}
\usepackage{indentfirst}
\usepackage{subscript}
\usepackage[multiple]{footmisc}

\usepackage{ifthen}
\usepackage{calc}
\usepackage{pifont}
\usepackage{color}
\usepackage{fancyhdr}
\usepackage{tikz}

\newcounter{hours}
\newcounter{minutes}

\newboolean{timeofmake}
\setboolean{timeofmake}{false}

\newcommand{\ignore}[1]{}

\newcommand{\intersub}[0]{{Subarray Access Refresh Parallelization}\xspace}

\newcommand{\is}[0]{{\text{SARP}}\xspace}
\newcommand{\sarp}[0]{{\text{SARP}}\xspace}
\newcommand{\isab}[0]{{$\text{SARP}_{\text{ab}}$}\xspace}
\newcommand{\ispb}[0]{{$\text{SARP}_{\text{pb}}$}\xspace}
\newcommand{\warplongCap}[0]{{\mbox{Write-refresh} Parallelization}\xspace}
\newcommand{\warplong}[0]{{\mbox{write-refresh} parallelization}\xspace}
\newcommand{\Warplong}[0]{{\mbox{Write-refresh} parallelization}\xspace}
\newcommand{\darp}[0]{{DARP}\xspace}
\newcommand{\darplong}[0]{{Dynamic Access Refresh Parallelization}\xspace}
\newcommand{\Ooolong}[0]{{\mbox{Out-of-order} per-bank refresh}\xspace}
\newcommand{\ooolong}[0]{{\mbox{out-of-order} per-bank refresh}\xspace}
\newcommand{\ooolongCap}[0]{{\mbox{Out-of-order} Per-bank Refresh}\xspace}
\newcommand{\ib}[0]{{\darp}\xspace}

\newcommand{\combo}[0]{{DSARP}\xspace}

\newcommand{\act}[0]{\texttt{\small{ACTIVATE}}\xspace}
\newcommand{\acts}[0]{\texttt{\small{ACTIVATEs}}\xspace}
\newcommand{\refab}[0]{{\small\emph{$REF_{ab}$}}\xspace}
\newcommand{\refpb}[0]{{\small\emph{$REF_{pb}$}}\xspace}
\newcommand{\trefi}[0]{{\small\emph{$tREFI_{ab}$}}\xspace}
\newcommand{\trefipb}[0]{{\small\emph{$tREFI_{pb}$}}\xspace}
\newcommand{\trfc}[0]{{\small\emph{$tRFC_{ab}$}}\xspace}
\newcommand{\trfcpb}[0]{{\small\emph{$tRFC_{pb}$}}\xspace}

\newcommand{\trrd}[0]{{\small\emph{tRRD}}\xspace}
\newcommand{\tfaw}[0]{{\small\emph{tFAW}}\xspace}
\newcommand{\fgr}[0]{{\small\emph{FGR}}\xspace}

\newcommand{\caprefab}[0]{\small{\textbf{\textit{REF\textsubscript{ab}}}}\xspace}
\newcommand{\caprefpb}[0]{\small{\textbf{\textit{REF\textsubscript{pb}}}}\xspace}
\newcommand{\captrfc}[0]{\small{\textbf{\textit{tRFC\textsubscript{ab}}}}\xspace}

\newcommand*\circled[1]{\tikz[baseline=(char.base)]{
            \node[shape=circle,draw,inner sep=0.8pt,fill=black,text=white] (char) {#1};}}

\newcommand{\figputHW}[3]{
\begin{figure}[h]
\begin{minipage}{\linewidth}
\footnotesize 
\begin{center}
\includegraphics[width=1.0\linewidth]{plots/#1}
\end{center}
\vspace{-0.15in}
\caption{#2 \label{fig:#1}}
\end{minipage}
\end{figure}
}
\newcommand{\figputHS}[3]{
\begin{figure}[h]
\begin{minipage}{\linewidth}
\begin{center}
\includegraphics[scale=#2]{plots/#1}
\end{center}
\vspace{-0.15in}
\caption{#3 \label{fig:#1}}
\end{minipage}
\end{figure}
}

\newcommand{\figputGHS}[3]{
\begin{figure}[h]
\begin{minipage}{\linewidth}
\begin{center}
\includegraphics[scale=#2]{gnuplots/#1}
\end{center}
\vspace{-0.15in}
\caption{#3 \label{fig:#1}}
\end{minipage}
\end{figure}
}

\newcommand{\figputGTS}[3]{
\begin{figure*}[t]
\begin{minipage}{\linewidth}
\begin{center}
\includegraphics[scale=#2]{gnuplots/#1}
\end{center}
\vspace{-0.15in}
\caption{#3 \label{fig:#1}}
\end{minipage}
\end{figure*}
}

\newcommand{\figref}[1]{Figure~\ref{fig:#1}}

\begin{document}
\title{Improving DRAM Performance by Parallelizing Refreshes with Accesses}

\preauthor{\begin{center}\large
    \begin{tabular}[t]{ccc}
          Kevin Kai-Wei Chang & Donghyuk Lee & Zeshan Chishti$\dagger$ \\
          \authemail{kevincha@cmu.edu} & \authemail{donghyu1@cmu.edu} &
          \authemail{zeshan.a.chishti@intel.com} \\
    \end{tabular}
        \vskip .5em
    \begin{tabular}[t]{cccc}
        Alaa R. Alameldeen$\dagger$ &  Chris Wilkerson$\dagger$ & Yoongu Kim & Onur Mutlu \\
        \authemail{alaa.r.alameldeen@intel.com} & \authemail{chris.wilkerson@intel.com} &
        \authemail{yoongukim@cmu.edu} & \authemail{onur@cmu.edu} \\
    \end{tabular}
        \vskip .5em
    \begin{tabular}[t]{cc}
        Carnegie Mellon University & $\dagger$Intel Labs \\
}
\postauthor{\end{tabular}\par\end{center}\vspace{-0.2in}}


\date{}
\maketitle

\thispagestyle{empty}
{
\setstretch{0.94}
\frenchspacing
\begin{abstract}




Modern DRAM cells are periodically refreshed to prevent data loss due
to leakage. Commodity DDR (double data rate) DRAM refreshes cells at
the rank level. This degrades performance significantly because it
prevents an entire DRAM rank from serving memory requests while being
refreshed. DRAM designed for mobile platforms, LPDDR (low power DDR)
DRAM, supports an enhanced mode, called per-bank refresh, that
refreshes cells at the bank level. This enables a bank to be accessed
while another in the same rank is being refreshed, alleviating part of
the negative performance impact of refreshes. Unfortunately, there are
two shortcomings of per-bank refresh employed in today's systems.
First, we observe that the per-bank refresh scheduling scheme does not
exploit the full potential of overlapping refreshes with accesses
across banks because it restricts the banks to be refreshed in a
sequential round-robin order. Second, accesses to a bank that is being
refreshed have to wait.





To mitigate the negative performance impact of DRAM refresh, we
propose two complementary mechanisms, \ib (\darplong) and \is
(\intersub). The goal is to address the drawbacks of per-bank refresh
by building more efficient techniques to parallelize refreshes and
accesses within DRAM. First, instead of issuing per-bank refreshes in a
round-robin order, as it is done today, \ib issues per-bank refreshes
to idle banks in an out-of-order manner. Furthermore, \ib proactively
schedules refreshes during intervals when a batch of writes are
draining to DRAM. Second, \is exploits the existence of mostly-independent
{\em subarrays} within a bank. With minor modifications to DRAM
organization, it allows a bank to serve memory accesses to an idle
subarray while another subarray is being refreshed. Extensive
evaluations on a wide variety of workloads and systems show that our
mechanisms improve system performance (and energy efficiency) compared
to three \mbox{state-of-the-art} refresh policies and the performance
benefit increases as DRAM density increases.

\end{abstract}

\section{Introduction} \label{sec:intro}

Modern main memory is predominantly built using \emph{dynamic random
  access memory} (DRAM) cells. A DRAM cell consists of a capacitor to
store one bit of data as electrical charge. The capacitor leaks charge
over time, causing stored data to change. As a result, DRAM requires
an operation called \emph{refresh} that periodically restores
electrical charge in DRAM cells to maintain data integrity.








Each DRAM cell must be refreshed periodically every {\em refresh interval} as
specified by the DRAM standards~\cite{jedec-ddr3,jedec-lpddr3}. The exact
refresh interval time depends on the DRAM type (e.g., DDR or LPDDR) and the
operating temperature.  While DRAM is being refreshed, it becomes unavailable to
serve memory requests. As a result, refresh latency significantly degrades
system performance~\cite{liu-isca2012, mukundan-isca2013,
nair-hpca2013,stuecheli-micro2010} by delaying in-flight memory requests. This
problem will become more prevalent as DRAM density increases, leading to more
DRAM rows to be refreshed within the same refresh interval.  DRAM chip density
is expected to increase from 8Gb to 32Gb by 2020 as it doubles every two to
three years~\cite{itrs-dram}. Our evaluations show that DRAM refresh, as it is
performed today, causes an average performance degradation of 8.2\% and 19.9\%
for 8Gb and 32Gb DRAM chips, respectively, on a variety of memory-intensive
workloads running on an 8-core system. Hence, it is important to develop
practical mechanisms to mitigate the performance penalty of DRAM refresh.




There are two major ways refresh operations are performed in modern DRAM
systems: {\em all-bank refresh (or, rank-level refresh)} and {\em per-bank
refresh}. These methods differ in what levels of the DRAM hierarchy refresh
operations tie up to. A modern DRAM system is organized as a hierarchy of ranks and
banks. Each rank is composed of multiple banks. Different ranks and banks can be
accessed independently. Each bank contains a number of rows (e.g., 16-32K in
modern chips). Because successively refreshing {\em all} rows in a DRAM chip
would cause very high delay by tying up
the entire DRAM device, modern memory controllers issue a number of refresh
commands that are evenly distributed throughout the refresh
interval~\cite{jedec-ddr3,jedec-lpddr3,liu-isca2013,liu-isca2012,nair-hpca2013}.
Each refresh command refreshes a small number of rows.\footnote{The time between
two refresh commands is fixed to an amount that is dependent on the DRAM type
and temperature.} The two common refresh methods of today differ in where in
the DRAM hierarchy the rows refreshed by a refresh command reside.

In {\em all-bank refresh}, employed by both commodity DDR and LPDDR
DRAM chips, a refresh command operates at the rank level: it refreshes
a number of rows in {\em all} banks of a rank concurrently. This
causes every bank within a rank to be unavailable to serve memory
requests until the refresh command is complete.  Therefore, it
degrades performance significantly, as we demonstrate in
Section~\ref{sec:motivation} and as others have
demonstrated~\cite{liu-isca2012, mukundan-isca2013,
  nair-hpca2013,stuecheli-micro2010}. In {\em per-bank refresh},
employed by LPDDR DRAM~\cite{jedec-lpddr3, micronLPDDR2_2Gb} as an
alternative refresh mode, a refresh command operates at the bank
level: it refreshes a number of rows in only a single bank of a rank
at a time.\footnote{One can think of per-bank refresh as splitting up
  a single large all-bank refresh operation performed on an entire
  rank into smaller groups of refresh operations performed on each
  bank.} This enables a bank to be accessed while another in the same
rank is being refreshed, alleviating part of the negative performance
impact of refreshes (as we show in
Section~\ref{sec:motivation}). Unfortunately, per-bank refresh suffers
from two shortcomings that limit the ability of DRAM to serve requests
while refresh operations are being performed. First, we observe that
the per-bank refresh scheduling scheme does not exploit the full
potential of overlapping refreshes with accesses across banks because
it restricts the banks to be refreshed in a strict sequential {\em
  round-robin order}~\cite{jedec-lpddr3}. Second, accesses to a bank
that is being refreshed still have to wait as there is no way today to
perform a refresh and an access to the same bank concurrently. We find
that, due to these shortcomings, a significant performance degradation
still exists with per-bank refresh (Section~\ref{sec:motivation}).

Our goal is to alleviate the shortcomings of per-bank refresh by enabling more
efficient parallelization of refreshes and accesses within DRAM. The major ideas
are to (1) provide a more flexible scheduling policy for traditional per-bank
refresh and (2) exploit the internal {\em subarray} structure of a bank to
enable parallelization of refreshes and accesses within a bank. To this end, we
propose two complementary techniques.


The first technique, \emph{\darplong (\darp)}, is a new refresh
scheduling policy based on two key ideas: \emph{\ooolong} and
\emph{\warplong}. The first idea enables the memory controller to
specify an idle bank to be refreshed as opposed to the
state-of-the-art per-bank refresh policy that refreshes banks in a
strict round-robin order. By monitoring the bank request queues'
occupancies, \darp avoids refreshing a bank that has pending memory
requests and instead refreshes an idle bank to maximize
parallelization of refreshes and accesses. The second idea hides
refresh latency with write accesses by proactively scheduling per-bank
refreshes during the time period when banks are serving write
requests. There are two reasons why we attempt to parallelize refresh
operations with write accesses. First, modern systems typically buffer
write requests in memory controllers and drain them to DRAM in a batch
to mitigate the \mbox{\emph{bus turnaround}}
penalty~\cite{chatterjee-hpca2012,lee-tech2010,stuecheli-isca2010}.
Write batching allows our mechanism to proactively schedule a per-bank
refresh to a bank while other banks are serving the accumulated write
requests, thus enabling more parallelization of refreshes and writes
to hide refresh latency.  Second, although \ib can potentially delay
write requests to some of the banks, this does not significantly
affect system performance. The reason is that DRAM writes (which are
writebacks from the last-level
cache~\cite{lee-tech2010,stuecheli-isca2010}) are not latency-critical
as processors do not stall to wait for them to finish.






The second technique, \emph{\intersub (\is)}, takes advantage of the
fact that a DRAM bank is composed of multiple \emph{subarrays} and
each subarray has its own local \emph{sense
  amplifiers}~\cite{itoh-vlsi, kim-isca2012, lee-hpca2013,
  moon-isscc2009,vogelsangmoon-micro2010}. We observe that only a few
subarrays are refreshed within a bank when the other subarrays and the
DRAM I/O bus remain completely idle. Based on this observation,
\emph{\is} enables a bank to be accessible while being refreshed: it
can serve read or write requests to idle subarrays while other
subarrays in the bank are being refreshed. Therefore, \is reduces the
interference of refreshes on demand requests at the cost of very
modest modifications to DRAM devices.

We make the following major \textbf{contributions}:

\begin{itemize}

\item We propose a new per-bank refresh scheduling policy, \darp
  (\darplong), to proactively schedule refreshes to banks that are
  idle or that are draining writes.

\item We propose a new DRAM refresh mechanism, \is (\intersub), to
  enable a bank to serve memory requests in idle subarrays while other
  subarrays are being refreshed.




\item We comprehensively evaluate the performance and energy benefits
  of \ib and \is, and their combination, \emph{\combo}, compared to
  three state-of-the-art refresh policies across a wide variety of
  workloads and system configurations.  One particular evaluation
  shows that \combo improves system performance by 3.3\%/7.2\%/15.2\%
  on average (and up to 7.1\%/14.5\%/27.0\%) across 100 workloads over
  the best previous mechanism (per-bank refresh) for 8/16/32Gb DRAM
  devices. \combo's performance gain increases as workload memory
  intensity and core count increase.



\end{itemize}

\section{Background}
\label{sec:background}

\subsection{DRAM System Organization}
\label{sec:dram-org}
At a high level, a DRAM system is
organized as a hierarchy of ranks and banks as shown in
\figref{dram-organization-v2}. Each rank consists of multiple banks that share
an internal bus for reading/writing data.\footnote{A typical DRAM system has 2
ranks connected to each channel and 8 banks per rank.} Because each bank acts as
an independent entity, banks can serve multiple memory requests in parallel,
offering \emph{bank-level
parallelism}~\cite{kim-micro2010,lee-micro2009,mutlu-isca2008}.

\figputHS{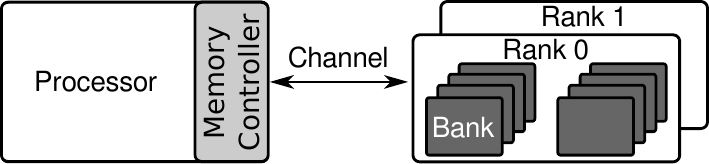}{1.0}{DRAM system organization.}

A DRAM bank is further sub-divided into multiple
\emph{subarrays}~\cite{kim-isca2012, seshadri-micro2013, vogelsangmoon-micro2010} as shown in
\figref{bank-organization}.
A subarray consists of a 2-D array of cells organized in rows and
columns.\footnote{Physically, DRAM has 32 to 64 subarrays, which
  varies depending on the number of rows (typically 16-32K) within a
  bank. We divide them into 8 subarray groups and refer to a subarray
  group as a subarray~\cite{kim-isca2012}.} Each DRAM cell has two
components: 1) a \emph{capacitor} that stores one bit of data as
electrical charge and 2) an \emph{access transistor} that connects the
capacitor to a wire called \emph{bitline} that is shared by a column
of cells. The access transistor is controlled by a wire called
\emph{wordline} that is shared by a row of cells.  When a wordline is
raised to $V_{DD}$, a row of cells becomes connected to the bitlines,
allowing reading or writing data to the connected row of cells. The
component that reads or writes a bit of data on a bitline is called a
\emph{sense amplifier}, shared by an entire column of cells. A row of
sense amplifiers is also called a \emph{row buffer}. All subarrays'
row buffers are connected to an I/O buffer~\cite{khoetal-isscc2009,
  moon-isscc2009} that reads and writes data from/to the bank's I/O
bus.



\figputHS{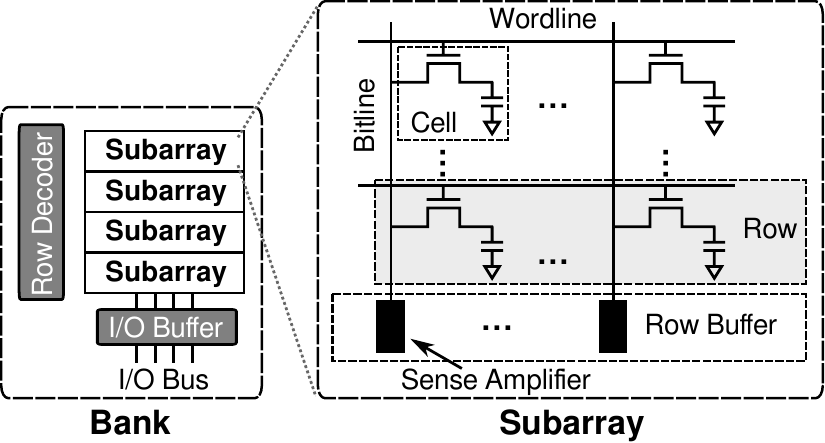}{0.8}{DRAM bank and subarray organization.}


\subsection{DRAM Refresh} 
\subsubsection{All-Bank Refresh (\refab)} The minimum time interval during which
any cell can retain its electrical charge without being refreshed is called the
\emph{minimum retention time}, which depends on the operating temperature and
DRAM type. Because there are tens of thousands of rows in DRAM, refreshing all
of them in bulk incurs high latency. Instead, memory controllers send a number
of refresh commands that are evenly distributed throughout the retention time to
trigger refresh operations, as shown in \figref{ab_timeline}. Because a typical
refresh command in a commodity DDR DRAM chip operates at an entire rank level,
it is also called an \emph{all-bank refresh} or \emph{\refab} for
short~\cite{jedec-ddr3,jedec-lpddr3,micronLPDDR2_2Gb}. The timeline shows that the time
between two \refab commands is specified by \trefi (e.g., 7.8$\mu$s for 64ms
retention time). Therefore, refreshing a rank requires $\sfrac{64ms}{7.8\mu
s}\approx8192$ refreshes and each operation refreshes exactly $\sfrac{1}{8192}$
of the rank's rows.


When a rank receives a refresh command, it sends the command to a DRAM-internal
refresh unit that selects which specific rows or banks to refresh. A \refab
command triggers the refresh unit to refresh a number of rows in every bank for
a period of time called \trfc (\figref{ab_timeline}). During \trfc, banks are
not refreshed simultaneously. Instead, refresh operations are staggered
(pipelined) across banks~\cite{mukundan-isca2013}. The main reason is that
refreshing every bank simultaneously would draw more current than what the power
delivery network can sustain, leading to potentially incorrect DRAM
operation~\cite{mukundan-isca2013, shevgoor-micro2013}. Because a \refab command
triggers refreshes on all the banks within a rank, the rank cannot process any
memory requests during \trfc, The length of \trfc is a function of the number of
rows to be refreshed.


\begin{figure}[h]
\centering
\subfloat[All-bank refresh (\refab) frequency and granularity.]{
    \begin{minipage}{\linewidth}
    \begin{center}
    \label{fig:ab_timeline}
    \includegraphics[width=\linewidth]{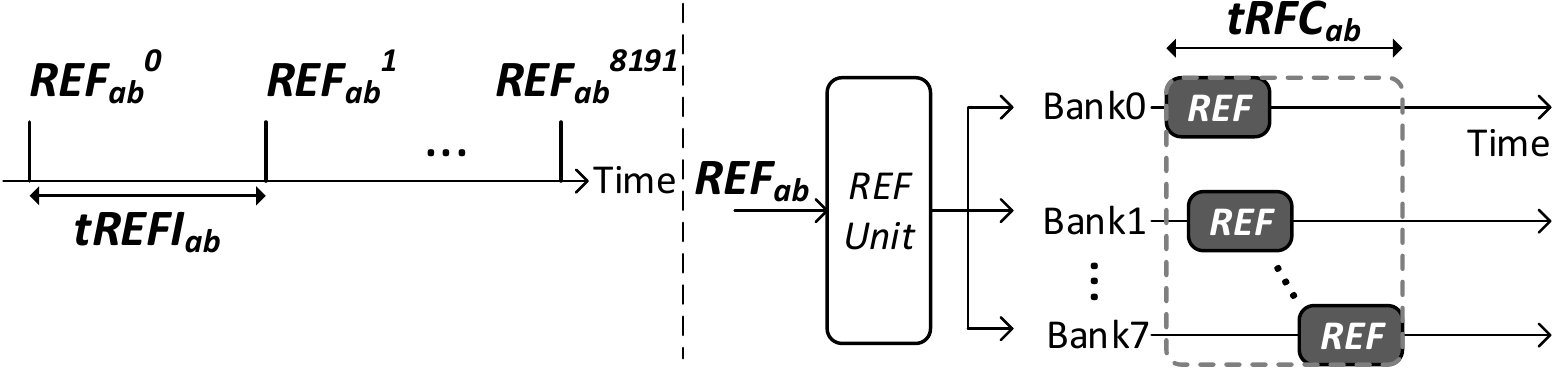}
    \end{center}
    \end{minipage}
}

\subfloat[Per-bank refresh (\refpb) frequency and granularity.]{
    \begin{minipage}{\linewidth}
    \begin{center}
    \label{fig:pb_timeline}
    \includegraphics[width=\linewidth]{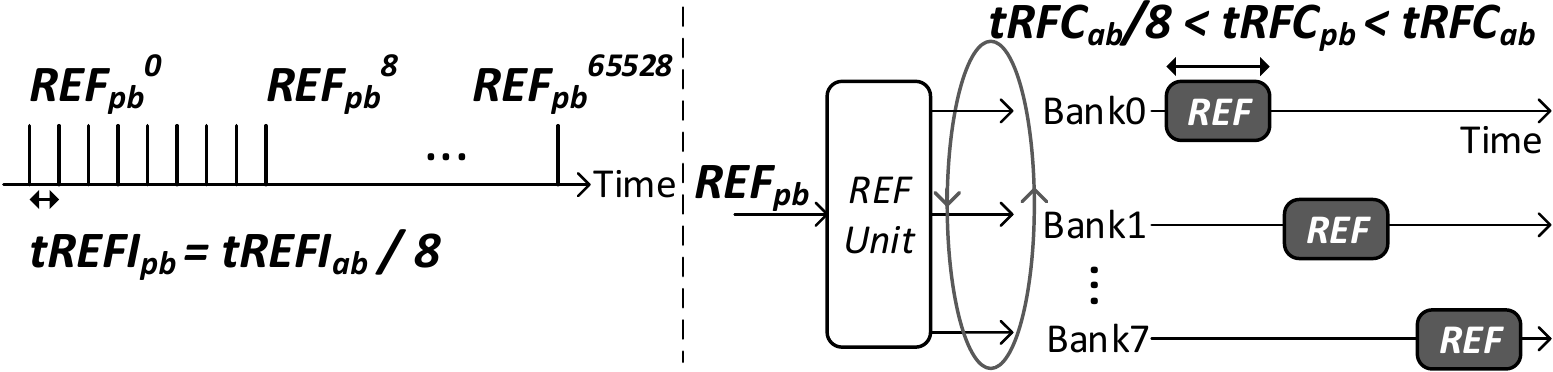}
    \end{center}
    \end{minipage}
}
\caption{Refresh command service timelines.}
\label{fig:inter-bank-service-timeline}
\end{figure}

\subsubsection{Per-Bank Refresh (\refpb)}
To allow partial access to DRAM during refresh, LPDDR DRAM (which is
designed for mobile platforms), supports an additional
finer-granularity refresh scheme, called \emph{per-bank refresh}
(\refpb for short)~\cite{jedec-lpddr3,micronLPDDR2_2Gb}. It splits up
a \refab operation into eight separate operations scattered across
eight banks (\figref{pb_timeline}).  Therefore, a \refpb command is
issued eight times more frequently than a \refab command (i.e.,
\trefipb = \trefi / 8).

Similar to issuing a \refab, a controller simply sends a \refpb
command to DRAM every \trefipb without specifying which particular
bank to refresh. Instead, when a rank's internal refresh unit receives
a \refpb command, it refreshes only one bank for each command
following a \emph{sequential round-robin order} as shown in
\figref{pb_timeline}. The refresh unit uses an internal counter to
keep track of which bank to refresh next.

By scattering refresh operations from \refab into multiple and
non-overlapping per-bank refresh operations, the refresh latency of
\refpb (\trfcpb) becomes shorter than \trfc. Disallowing \refpb
operations from overlapping with each other is a design decision made
by the LPDDR DRAM standard committee~\cite{jedec-lpddr3}. The reason
is simplicity: to avoid the need to introduce new timing constraints,
such as the timing between two overlapped refresh
operations.\footnote{At slightly increased complexity, one can
  potentially propose a modified standard that allows overlapped
  refresh of a subset of banks within a rank.}

With the support of \refpb, LPDDR DRAM can serve memory requests to
non-refreshing banks in parallel with a refresh operation in a single
bank.  \figref{per-bank-refresh-timeline} shows pictorially how \refpb
provides performance benefits over \refab from parallelization of
refreshes and reads.  \refpb reduces refresh interference on reads by
issuing a refresh to Bank 0 while Bank 1 is serving
reads. Subsequently, it refreshes Bank 1 to allow Bank 0 to serve a
read. As a result, \refpb alleviates part of the performance loss due
to refreshes by enabling parallelization of refreshes and accesses
across banks.



\begin{figure}[h]
\begin{center}
\includegraphics[width=\linewidth]{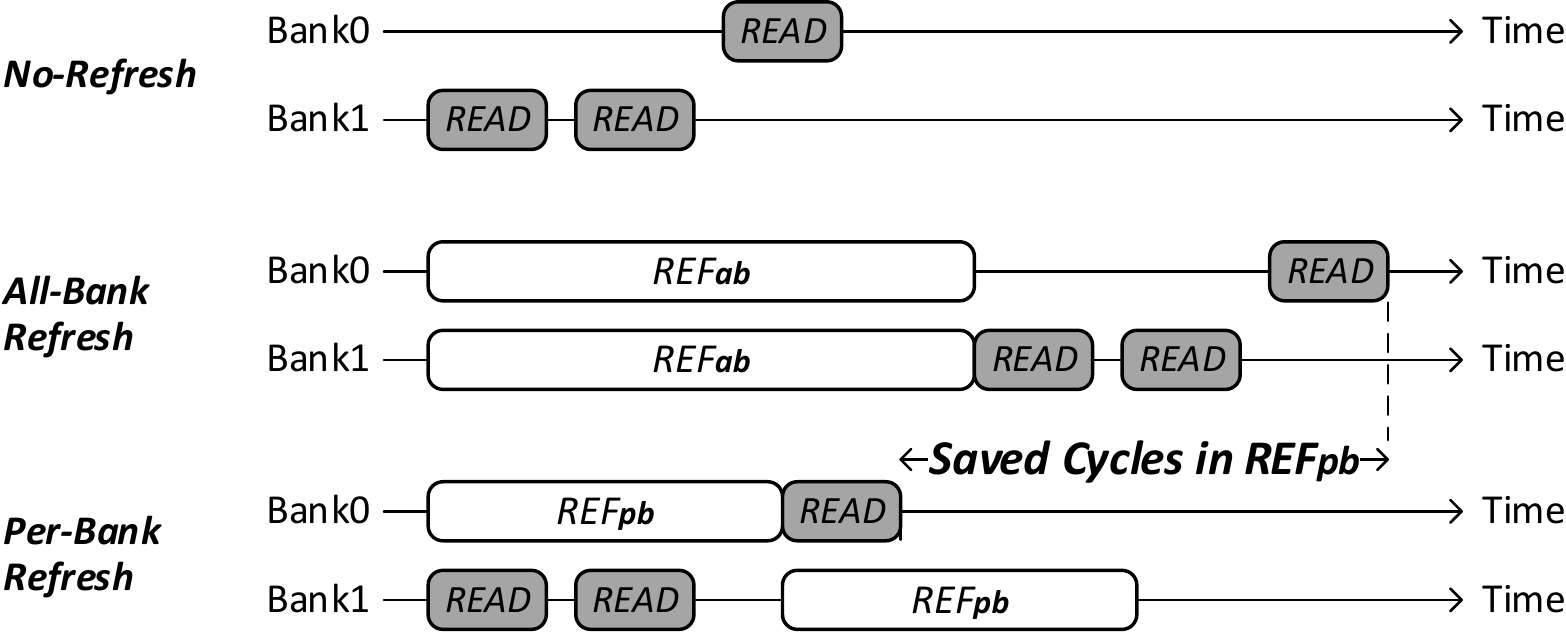}
\end{center}
\caption{Service timelines of all-bank and per-bank refresh.}
\label{fig:per-bank-refresh-timeline}
\end{figure}

\section{Motivation}
\label{sec:motivation}


In this section, we first describe the scaling trend of commonly used
all-bank refresh in both LPDDR and DDR DRAM as chip density increases
in the future. We then provide a quantitative analysis of all-bank
refresh to show its performance impact on multi-core systems followed
by performance comparisons to per-bank refresh that is only supported
in LPDDR.

\subsection{Increasing Performance Impact of Refresh}
\label{sec:refresh-challenge}


During the \trfc time period, the entire memory rank is locked up, preventing
the memory controller from sending any memory request. As a result, refresh
operations degrade system performance by increasing the latency of memory
accesses. The negative impact on system performance is expected to be
exacerbated as \trfc increases with higher DRAM density. The value of \trfc is
currently 350ns for an 8Gb memory device~\cite{jedec-ddr3}. \figref{rfc_trend}
shows our estimated trend of \trfc for future DRAM generations using linear
extrapolation on the currently available and previous DRAM devices. The same
methodology is used in prior works~\cite{liu-isca2012,stuecheli-micro2010}.
\emph{Projection 1} is an extrapolation based on 1, 2, and 4Gb devices;
\emph{Projection 2} is based on 4 and 8Gb devices. We use the more optimistic
\emph{Projection 2} for our evaluations. As it shows, \trfc may reach up to
1.6$\mu$s for future 64Gb DRAM devices. This long period of unavailability to
process memory accesses is detrimental to system performance.

\figputGHS{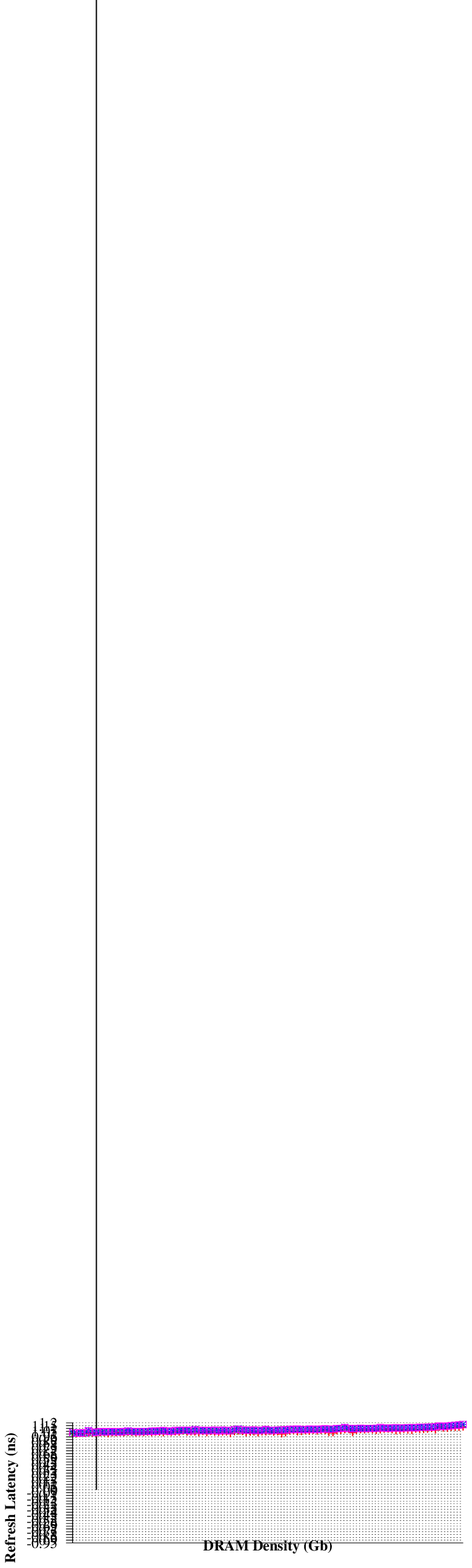}{1.0}{Refresh latency (\captrfc) trend.}

To demonstrate the negative system performance impact of DRAM refresh,
we evaluate 100 randomly mixed workloads categorized to five different
groups based on memory intensity on an 8-core system using various
DRAM densities.\footnote{Detailed methodology is described in
  Section~\ref{methodology}, including workloads, simulation
  methodology, and performance metrics.\label{fnote}} We use up to
32Gb DRAM density that the ITRS predicts to be manufactured by
2020~\cite{itrs-dram}. \figref{perf_loss_mpki} shows the average
performance loss due to all-bank refresh compared to an ideal baseline
without any refreshes for each memory-intensity category. The
performance degradation due to refresh becomes more severe as either
DRAM chip density (i.e., \trfc) or workload memory intensity increases
(both of which are trends in systems), demonstrating that it is
increasingly important to address the problem of DRAM refresh.

\figputGHS{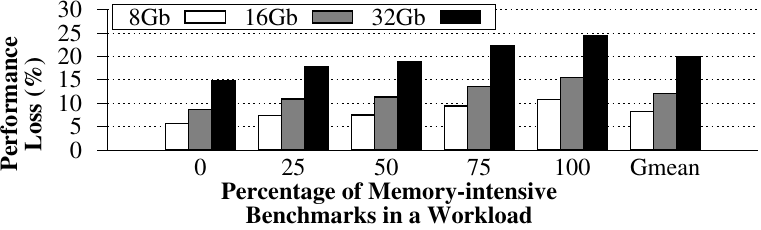}{1.0}{Performance degradation due to refresh.}



Even though the current DDR3 standard does not support \refpb, we believe that
it is important to evaluate the performance impact of \refpb on DDR3 DRAM
because DDR3 DRAM chips are widely deployed in desktops and servers.
Furthermore, adding per-bank refresh support to a DDR3 DRAM chip should be
non-intrusive because it does not change the internal bank organization. We
estimate the refresh latency of \refpb in a DDR3 chip based on the values used
in an LPDDR2 chip. In a 2Gb LPDDR2 chip, the per-bank refresh latency (\trfcpb)
is 90ns and the all-bank refresh latency (\trfc) is 210ns, which takes
\emph{2.3x} longer than \trfcpb~\cite{micronLPDDR2_2Gb}.\footnote{LPDDR2 has a
shorter \trfc than DDR3 because LPDDR2 1) has a retention time of 32ms instead
of 64ms in DDR3 under normal operating temperature and 2) each operation
refreshes fewer rows.} We apply this multiplicative factor to \trfc to calculate
\trfcpb.

Based on the estimated \trfcpb values, we evaluate the performance impact of
\refpb on the same 8-core system and workloads.\textsuperscript{\ref{fnote}}
\figref{perf_loss_all} shows the average performance degradation of \refab and
\refpb compared to an ideal baseline without any refreshes. Even though \refpb
provides performance gains over \refab by allowing DRAM accesses to
non-refreshing banks, its performance degradation becomes exacerbated as \trfcpb
increases with higher DRAM density. With 32Gb DRAM chips using \refpb, the
performance loss due to DRAM refresh is still a significant 16.6\% on average,
which motivates us to address issues related to \refpb.


\figputGHS{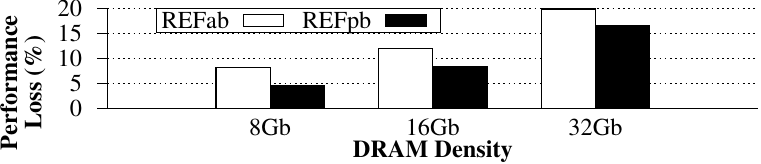}{1.0}{Performance loss due to \caprefab and \caprefpb.}


\subsection{Our Goal} We identify two main problems that \refpb faces. First,
\refpb commands are scheduled in a very restrictive manner in today's systems.
Memory controllers have to send \refpb commands in a sequential round-robin
order without any flexibility. Therefore, the current implementation does not
exploit the full benefit from overlapping refreshes with accesses across banks.
Second, \refpb cannot serve accesses to a refreshing bank until the refresh of
that bank is complete.  Our goal is to provide practical mechanisms to address
these two problems so that we can minimize the performance overhead of DRAM
refresh. 




%

\section{Mechanisms}
\label{sec:mechanism}

\subsection{Overview}


We propose two mechanisms, \emph{\darplong (\darp)} and
\emph{\intersub (\is)}, that hide refresh latency by parallelizing
refreshes with memory accesses across \emph{banks} and
\emph{subarrays}, respectively. \darp is a new refresh scheduling
policy that consists of two components. The first component is
\emph{\ooolong} that enables the memory controller to specify a
particular (idle) bank to be refreshed as opposed to the standard
per-bank refresh policy that refreshes banks in a strict round-robin
order. With out-of-order refresh scheduling, \darp can avoid
refreshing (non-idle) banks with pending memory requests, thereby
avoiding the refresh latency for those requests. The second component
is \emph{\warplong} that proactively issues per-bank refresh to a bank
while DRAM is draining write batches to other banks, thereby
overlapping refresh latency with write latency. The second mechanism,
\is, allows a bank to serve memory accesses in idle subarrays while
other subarrays within the same bank are being refreshed. \is exploits
the fact that refreshing a row is contained within a subarray, without
affecting the other subarrays' components and the I/O bus used for
transferring data. We now describe each mechanism in detail.





\subsection{\darplong}

\subsubsection{\ooolongCap} The limitation of the current per-bank refresh
mechanism is that it disallows a memory controller from specifying
which bank to refresh. Instead, a DRAM chip has internal logic that
strictly refreshes banks in a \emph{sequential round-robin order}.
Because DRAM lacks visibility into a memory controller's state (e.g.,
request queues' occupancy), simply using an in-order \refpb policy can
unnecessarily refresh a bank that has multiple pending memory requests
to be served when other banks may be free to serve a refresh
command. To address this problem, we propose the first component of
\darp, \emph{\ooolong}. The idea is to remove the bank selection logic
from DRAM and make it the memory controller's responsibility to
determine which bank to refresh. As a result, the memory controller
can refresh an idle bank to enhance parallelization of refreshes and
accesses, avoiding refreshing a bank that has pending memory requests
as much as possible.

Due to \refpb reordering, the memory controller needs to guarantee
that deviating from the original in-order schedule still preserves
data integrity. To achieve this, we take advantage of the fact that
the contemporary DDR JEDEC standard~\cite{jedec-ddr3,jedec-ddr4}
actually provides some refresh scheduling flexibility. The standard
allows up to \emph{eight} all-bank refresh commands to be issued late
(postponed) or early (pulled-in). This implies that each bank can
tolerate up to eight \refpb to be postponed or pulled-in. Therefore,
the memory controller ensures that reordering \refpb preserves data
integrity by limiting the number of postponed or pulled-in commands.

\figref{darp_flow_chart} shows the algorithm of our mechanism. The \ooolong
scheduler makes a refresh decision every DRAM cycle. There are three key steps.
First, when the memory controller hits a per-bank refresh schedule time (every
\trefipb), it postpones the scheduled \refpb if the to-be-refreshed bank
(\texttt{\emph{R}}) has pending demand requests (read or write) {\em and} it has
postponed fewer refreshes than the limit of eight (\circled{1}). The hardware
counter that is used to keep track of whether or not a refresh can be postponed
for each bank is called the \emph{refresh credit (ref\_credit)}. The counter
decrements on a postponed refresh and increments on a pulled-in refresh for each
bank.
Therefore, a \refpb command can be postponed if the bank's ref\_credit
stays above -8. Otherwise the memory controller is required to send a REFpb
command to comply with the standard.
Second, the memory controller prioritizes issuing commands for a demand request
if a refresh is not sent at any given time (\circled{2}). Third, if the memory
controller cannot issue any commands for demand requests due to the timing
constraints, it instead randomly selects one bank (\texttt{\emph{B}}) from a
list of banks that have no pending demand requests to refresh. Such a refresh
command is either a previously postponed \refpb or a new pulled-in \refpb
(\circled{3}).



\figputHS{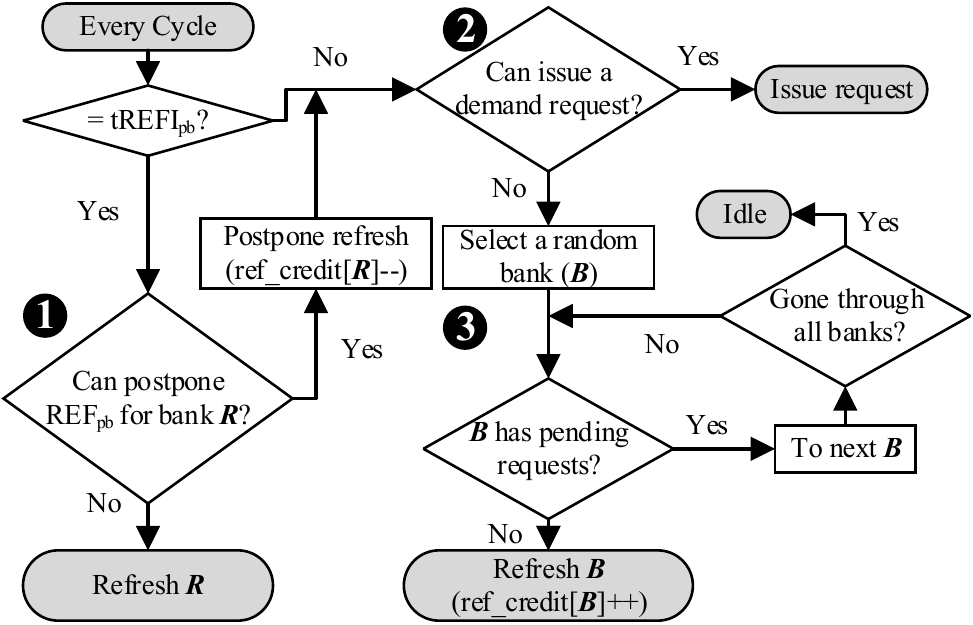}{0.85}{Algorithm of \ooolong.}

\subsubsection{\warplongCap} The key idea of the second component of \darp is to
actively avoid refresh interference on read requests and instead enable more
parallelization of refreshes with \emph{write requests}. We make two
observations that lead to our idea. First, {\em write batching} in DRAM creates
an opportunity to overlap a refresh operation with a sequence of writes, without
interfering with reads. A modern memory controller typically buffers DRAM writes
and drains them to DRAM in a batch to amortize the \emph{bus turnaround
latency}, also called \emph{tWTR} or
\emph{tRTW}~\cite{jedec-ddr3,kim-isca2012,lee-tech2010}, which is the additional
latency incurred from switching between serving writes to reads because DRAM I/O
bus is half-duplex. Typical systems start draining writes when the write buffer
occupancy exceeds a certain threshold until the buffer reaches a low watermark.
This draining time period is called the \emph{writeback mode}, during which no
rank within the draining channel can serve read
requests~\cite{chatterjee-hpca2012,lee-tech2010,stuecheli-isca2010}. Second,
DRAM writes are not latency-critical because processors do not stall to wait for
them: DRAM writes are due to dirty cache line evictions from the last-level
cache~\cite{lee-tech2010,stuecheli-isca2010}.

Given that writes are not latency-critical and are drained in a batch
for some time interval, we propose the second component of \darp,
\emph{\warplong}, that attempts to maximize parallelization of
refreshes and writes. \Warplong selects the bank with the minimum
number of pending demand requests (both read and write) and preempts
the bank's writes with a per-bank refresh. As a result, the bank's
refresh operation is hidden by the writes in other banks.


The reasons why we select the bank with the lowest number of demand
requests as a refresh candidate during writeback mode are
two-fold. First, the goal of the writeback mode is to drain writes as
fast as possible to reach a low watermark that determines the end of
the writeback
mode~\cite{chatterjee-hpca2012,lee-tech2010,stuecheli-isca2010}. Extra
time delay on writes can potentially elongate the writeback mode by
increasing queueing delay and reducing the number of writes served in
parallel across banks. Refreshing the bank with the lowest write
request count (zero or more) has the smallest impact on the writeback
mode length because other banks can continue serving their writes to
reach to the low watermark. Second, if the refresh scheduled to a bank
during the writeback mode happens to extend beyond writeback mode, it
is likely that the refresh 1) does not delay immediate reads within
the same bank because the selected bank has no reads or 2) delays
reads in a bank that has less contention. Note that we only preempt
one bank for refresh because the JEDEC standard~\cite{jedec-lpddr3}
disallows overlapping per-bank refresh operations across banks within
a rank.





\figref{inter-bank-service-timeline} shows the service timeline and
benefits of \warplong. There are \textbf{two scenarios} when the
scheduling policy parallelizes refreshes with writes to increase
DRAM's availability to serve read
requests. \figref{inter-bank-postpone} shows the first scenario when
the scheduler \emph{postpones} issuing a \refpb command to avoid
delaying a read request in Bank 0 and instead serves the refresh in
parallel with writes from Bank 1, effectively hiding the refresh
latency in the writeback mode. Even though the refresh can potentially
delay individual write requests during writeback mode, the delay does
not impact performance as long as the length of writeback mode remains
the same as in the baseline due to longer prioritized write request
streams in other banks. In the second scenario shown in
\figref{inter-bank-pull}, the scheduler proactively \emph{pulls in} a
\refpb command early in Bank 0 to fully hide the refresh latency from
the later read request while Bank 1 is draining writes during the
writeback mode (note that the read request cannot be scheduled during
the writeback mode).

\begin{figure}[t]
\centering
\subfloat[Scenario 1: Parallelize postponed refresh with writes.]{
    \begin{minipage}{\linewidth}
    \begin{center}
    \label{fig:inter-bank-postpone}
    \includegraphics[width=\linewidth]{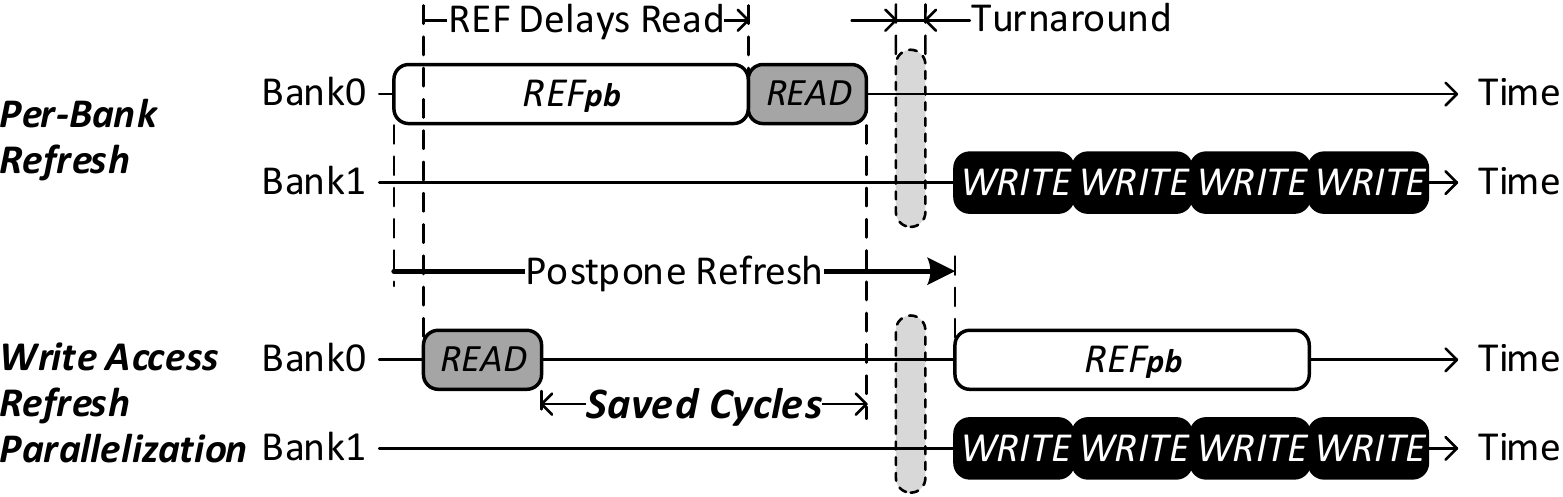}
    \end{center}
    \end{minipage}
}

\subfloat[Scenario 2: Parallelize pulled-in refresh with writes.]{
    \begin{minipage}{\linewidth}
    \begin{center}
    \label{fig:inter-bank-pull}
    \includegraphics[width=\linewidth]{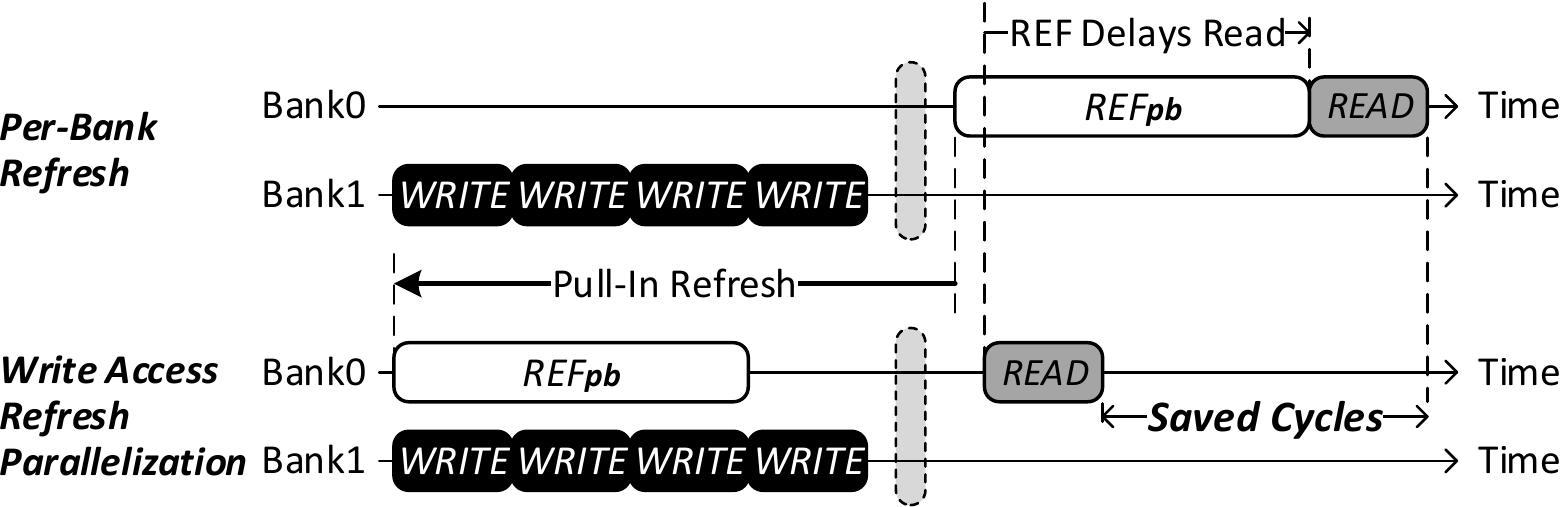}
    \end{center}
    \end{minipage}
}
\caption{Service timeline of a per-bank refresh operation along with read and
write requests using different scheduling policies.}
\label{fig:inter-bank-service-timeline}
\end{figure}

The crucial observation is that \warplong improves performance because
it avoids stalling the read requests due to refreshes by postponing or
pulling in refreshes in parallel with writes without extending the
writeback period.

\ignore{\figref{inter-bank-service-timeline} shows the service timeline and benefits of
\ib. There are \textbf{two scenarios} when the scheduling policy parallelizes
refresh operations with writes to increase DRAM's availability to serve read
requests. The first scenario is when the scheduler \emph{postpones} issuing a
\refpb to allow a DRAM bank to serve read requests. Then when the DRAM enters
writeback mode issuing write requests to other banks, the scheduler can send
the postponed refresh to the bank. As a result, the refresh operation is
parallelized with writes, effectively hiding the refresh latency in the
writeback mode to reduce the negative impact of refresh on read requests. Even
though refresh can potentially delay individual write requests during
writeback mode, the delay may not impact performance as long as the length
of writeback mode remains approximately the same due to longer prioritized
write request streams in other banks. \figref{inter-bank-postpone} shows the
performance benefit of postponing a refresh and overlapping it with writes.

In the second scenario, the scheduler takes preventive action by issuing a \refpb
command early when there is an opportunity to fully hide the refresh latency
during the writeback mode. As shown in \figref{inter-bank-pull}, the scheduler
speculatively \emph{pulls in} a \refpb command during the writeback mode before
it reaches its refresh schedule in order to free up an upcoming refresh
interval to serve a read request.}

Algorithm~\ref{algo:warp-algo} shows the operation of \warplong. When
the memory controller enters the writeback mode, the scheduler selects
a bank candidate for refresh when there is no pending refresh. A bank
is selected for refresh under the following criteria: 1) the bank has
the lowest number of demand requests among all banks and 2) its
refresh credit has not exceeded the maximum \emph{pulled-in} refresh
threshold. After a bank is selected for refresh, its credit increments
by one to allow an additional refresh postponement.

\begin{algorithm}
\caption{\small Write-refresh parallelization}
\label{algo:warp-algo}
\footnotesize{
\textbf{\underline{Every \trfcpb in Writeback Mode:}}
\begin{algorithmic}
\IF{\textit{refresh\_queue[0:N-1].isEmpty()}}
    \STATE \textit{b = find\_bank\_with\_lowest\_request\_queue\_count AND $ref\_credit<8$}
        \STATE \textit{refreshBank($b$)}
        \STATE \textit{ref\_credit[$b$]} += 1
\ENDIF
\end{algorithmic}
}
\end{algorithm}




\subsubsection{Implementation}
\darp incurs a small overhead in the memory controller and DRAM
without affecting the DRAM cell array organization. There are five
main modifications. First, each refresh credit is implemented with a
hardware integer counter that either increments or decrements by up to
eight when a refresh command is pulled-in or postponed,
respectively. Thus, the storage overhead is very modest with 4 bits
per bank (32 bits per rank). Second, \darp requires logic to monitor
the status of various existing queues and schedule refreshes as
described. Despite reordering refresh commands, all DRAM timing
constraints are followed, notably \trrd and \trfcpb that limit when
\refpb can be issued to DRAM.  Third, the DRAM command decoder needs
modification to decode the bank ID that is sent on the address bus
with the \refpb command.  Fourth, the refresh logic that is located
outside of the banks and arrays needs to be modified to take in the
specified bank ID. Fifth, each bank requires a separate row counter to
keep track of which rows to refresh as the number of postponed or
pulled-in refresh commands differs across banks.  Our proposal limits
the modification to the least invasive part of the DRAM without
changing the structure of the dense arrays that consume the majority
of the chip area.

\subsection{\intersub} Even though \darp allows refreshes and accesses to occur
in parallel across different banks, \darp cannot deal with their collision
\emph{within a bank}. To tackle this problem, we propose \emph{\sarp
(\intersub)} that exploits the existence of subarrays within a bank. The key
observation leading to our second mechanism is that refresh occupies only a few
\emph{subarrays} within a bank whereas the other \emph{subarrays} and the
\emph{I/O bus} remain idle during the process of refreshing. The reasons for
this are two-fold.  First, refreshing a row requires only its subarray's sense
amplifiers that restore the charge in the row without transferring any data
through the I/O bus.  Second, each subarray has its own set of \emph{sense
amplifiers} that are not shared with other subarrays.

Based on this observation, SARP's key idea is to allow memory accesses
to an \emph{idle} subarray while another subarray is refreshing.
\figref{subarray-service-timeline-fig} shows the service timeline and the
performance benefit of our mechanism. As shown, \is reduces the read
latency by performing the read operation to Subarray 1 in parallel
with the refresh in Subarray 0. Compared to \ib, \is provides the
following advantages: 1) \is is applicable to both all-bank and
per-bank refresh, 2) \is enables memory accesses to a refreshing bank,
which cannot be achieved with \ib, and 3) \is also utilizes bank-level
parallelism by serving memory requests from multiple banks while the
entire rank is under refresh. \is requires modifications to 1) the
DRAM architecture because two distinct wordlines in different
subarrays need to be raised simultaneously, which cannot be done in
today's DRAM due to the shared peripheral logic among subarrays, 2)
the memory controller such that it can keep track of which subarray is
under refresh in order to send the appropriate memory request to an
idle subarray.

\figputHW{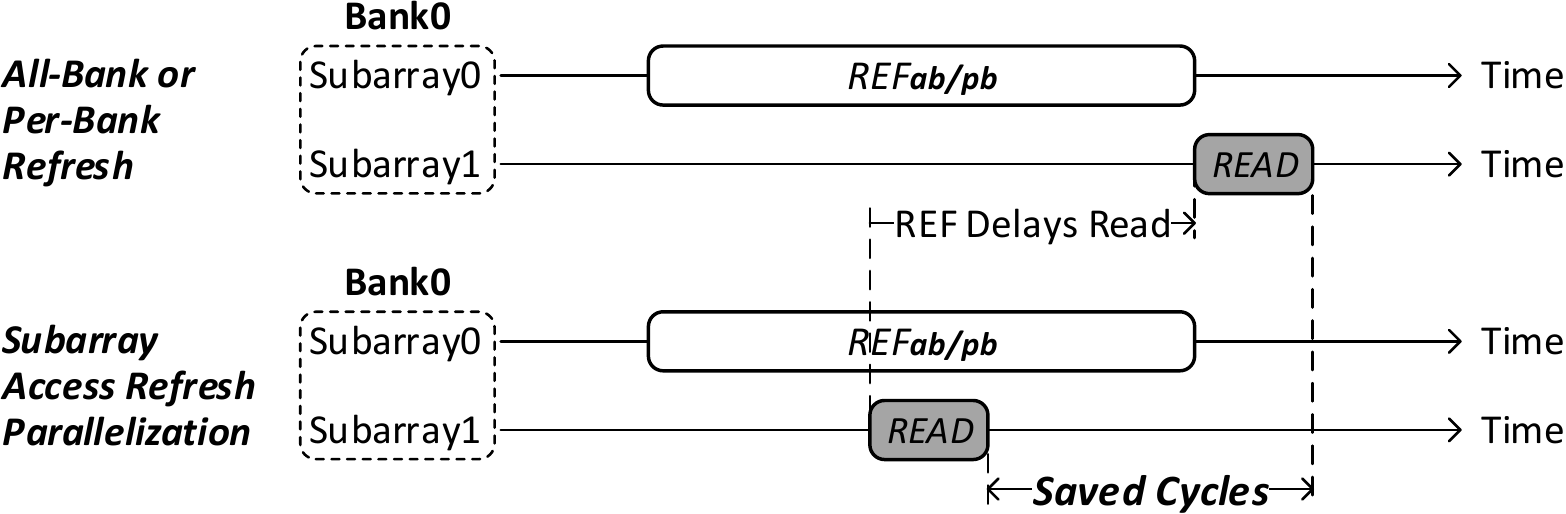}{Service timeline of a refresh and a read request to two
different subarrays within the same bank.}


\subsubsection{DRAM Bank Implementation for \is} \label{sec:sarp_implementation}
As opposed to \ib, \is requires modifications to DRAM to support accessing
subarrays individually. While subarrays are equipped with dedicated local
peripheral logic, what prevents the subarrays from being operated independently
is the global peripheral logic that is shared by all subarrays within a bank.


\figref{detail-bank-organization} shows a detailed view of an existing DRAM
bank's organization. There are two major shared peripheral components within a
bank that prevent modern DRAM chips to refresh at subarray level. First, each
bank has a \emph{global row decoder} that decodes the incoming row's addresses.
To read or write a row, memory controllers first issue an \act command with the
row's address.
Upon receiving this command, the bank feeds the row address to the \emph{global
row decoder} that broadcasts the partially decoded address to all subarrays
within the bank. After further decoding, the row's subarray then raises its
wordline to begin transferring the row's cells' content to the row
buffer.\footnote{The detailed step-to-step explanation of the activation process
can be found in prior
works~\cite{kim-isca2012,lee-hpca2013,seshadri-micro2013}.} During the transfer,
the row buffer also restores the charge in the row. Similar to an \act,
refreshing a row requires the refresh unit to \act the row to restore its
electrical charge (only the refresh row counter is shown for clarity in
\figref{detail-bank-organization}).
Because a bank has only one global row decoder and one pair of address wires
(for subarray row address and ID), it cannot simultaneously activate two
different rows (one for a memory access and the other for a refresh).

Second, when the memory controller sends a read or write command, the required
column from the activated row is routed through the \emph{global bitlines} into
the \emph{global I/O buffer} (both of which are shared across all subarrays' row
buffers) and is transferred to the I/O bus (all shown in
\figref{bank-organization}). This is done by asserting a \emph{column select}
signal that is routed globally to {\em all} subarrays, which enables {\em all}
subarrays' row buffers to be concurrently connected to the global bitlines.
Since this signal connects all subarrays' row buffers to the global bitlines at
the same time, if more than one activated row buffer (i.e., activated subarray)
exists in the bank, an electrical short-circuit occurs, leading to incorrect
operation. As a result, two subarrays cannot be kept activated when one is being
read or written to, which prevents a refresh to one subarray from happening
concurrently with an access in a different subarray in today's DRAM.

\begin{figure}[t]
\centering
\subfloat[Existing organization without SARP.]{
        \begin{minipage}{\linewidth}
    \begin{center}
    \label{fig:detail-bank-organization}
    \includegraphics[scale=0.7]{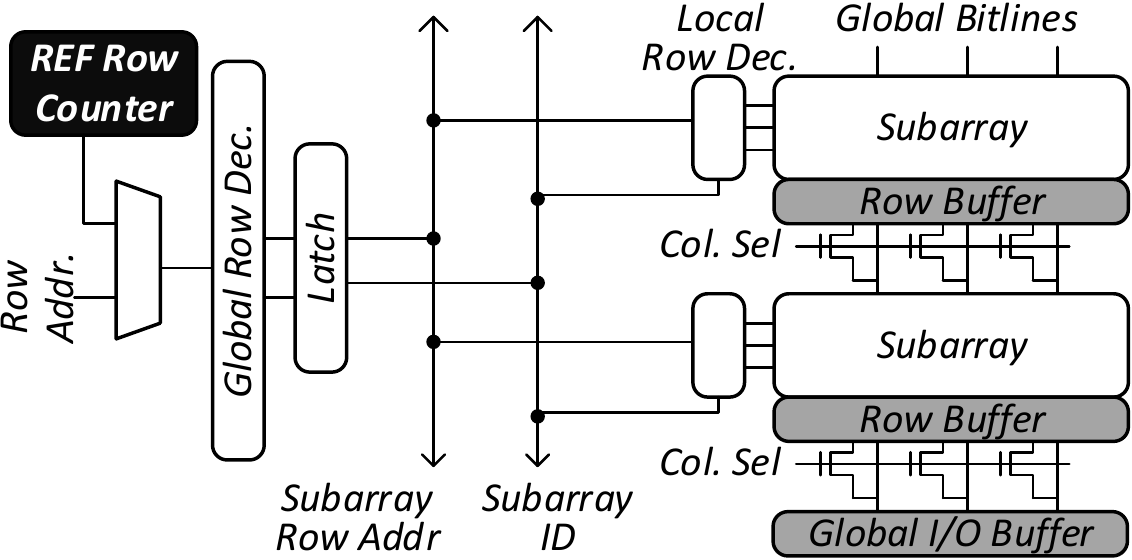}
    \end{center}
    \end{minipage}
}
\vspace{-0.1in}
\subfloat[New organization with SARP.]{
        \begin{minipage}{\linewidth}
    \begin{center}
    \label{fig:inter-sub-implementation}
    \includegraphics[scale=0.7]{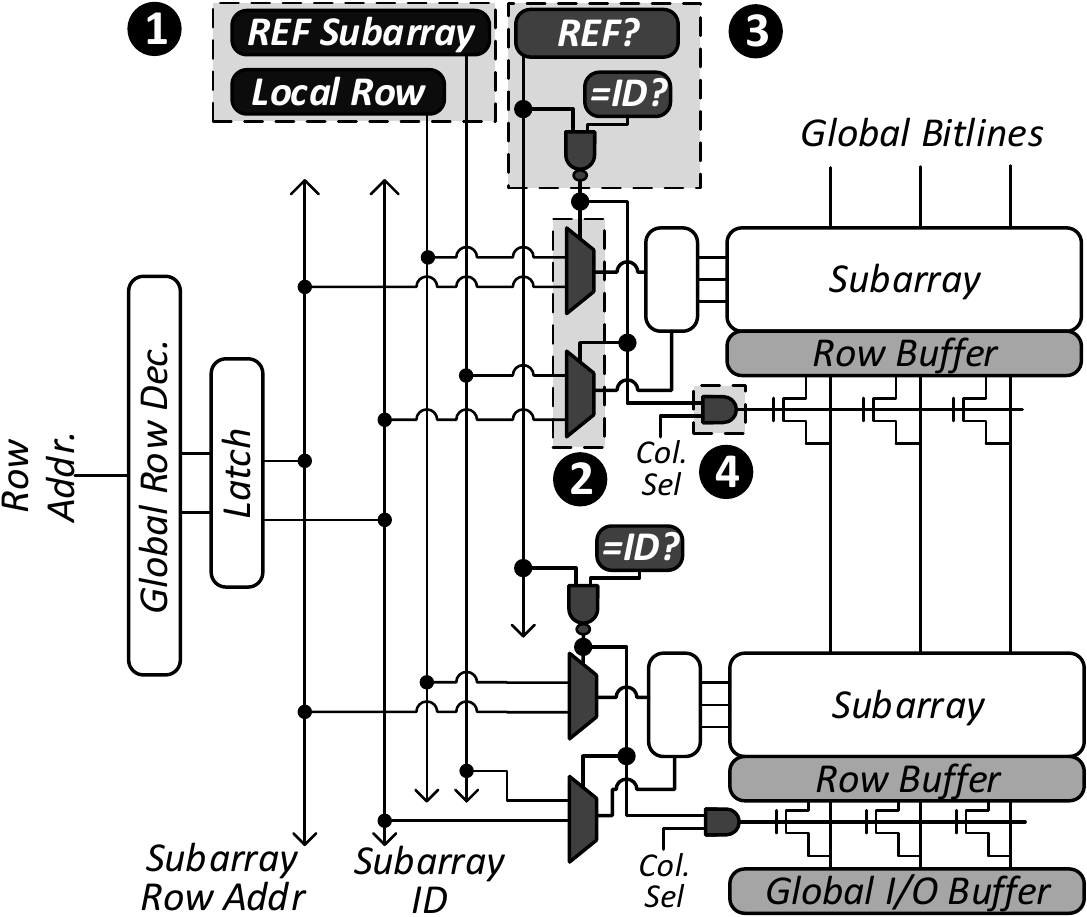}
    \end{center}
    \end{minipage}
}
\caption{DRAM bank without and with SARP.}
\end{figure}

The key idea of \is is to allow the concurrent activation of multiple subarrays,
but to only connect the accessed subarray's row buffer to the global bitlines
while another subarray is refreshing.\footnote{As described in
Section~\ref{sec:dram-org}, we refer to a subarray as a collection of multiple
physical subarrays. A modern refresh operation concurrently refreshes every
physical subarray within a collection.}  \figref{inter-sub-implementation} shows
our proposed changes to the DRAM microarchitecture. There are two major enablers
of \is.


The first enabler of \is allows both refresh and access commands to
simultaneously select their designated rows and subarrays with three
new components. The first component (\circled{1}) provides the subarray and
row addresses for refreshes without relying on the global row
decoder. To achieve this, it decouples the refresh row counter into a
\emph{refresh-subarray} counter and a \emph{local-row} counter that
keep track of the currently refreshing subarray and the row address
within that subarray, respectively. The second component (\circled{2})
allows each subarray to activate a row for either a refresh or an
access through two muxes. One mux is a row-address selector and the
other one is a subarray selector. The third component (\circled{3})
serves as a control unit that chooses a subarray for refresh. The
\texttt{\small{REF?}} block indicates if the bank is currently under
refresh and the \texttt{\small{=ID?}} comparator determines if the
corresponding subarray's ID matches with the refreshing subarray
counter for refresh. These three components form a new address path
for the refresh unit to supply refresh addresses in parallel with
addresses for memory accesses.

The second enabler of \is allows accesses to one activated subarray while
another subarray is kept activated for refreshes. We add an \emph{AND} gate
(\circled{4}) to each subarray that ensures the refreshing subarray's row buffer
is {\em not} connected to the global bitlines when the {\em column select}
signal is asserted on an access. At any instance, there is at most one activated
subarray among all non-refreshing subarrays because the global row decoder
activates only one subarray at a time. With the two proposed enablers, \is
allows one activated subarray for refreshes in parallel with another activated
subarray that serves data to the global bitlines.

\subsubsection{Detecting Subarray Conflicts in the Memory Controller} To avoid
accessing a refreshing subarray, which is determined internally by the DRAM chip
in our current mechanism, the memory controller needs to know the current
refreshing subarray and the number of subarrays. We create shadow copies of the
\emph{refresh-subarray} and \emph{local-row} counters in the memory controller
to keep track of the currently-refreshing subarray.
We store the number of subarrays in an EEPROM called the \emph{serial presence
detect (SPD)}~\cite{jedec-spd}, which stores various timing and DRAM
organization information in existing DRAM modules. The memory controller reads
this information at system boot time so that it can issue commands
correctly.\footnote{Note that it is possible to extend our mechanisms such that
the memory controller specifies the subarray to be refreshed instead of the DRAM
chip. This requires changes to the DRAM interface.}




\subsubsection{Power Integrity} Because an \act draws a lot of current, DRAM
standards define two timing parameters to constrain the activity rate
of DRAM so that \acts do not over-stress the power delivery
network~\cite{jedec-ddr3,shevgoor-micro2013}.  The first parameter is
the \emph{row-to-row activation delay} (\trrd) that specifies the
minimum waiting time between two subsequent \act commands within a
DRAM device. The second is called the \emph{four activate window}
(\tfaw) that defines the length of a rolling window during which a
maximum of four \acts can be in progress. Because a refresh operation
requires activating rows to restore charge in DRAM cells, \is consumes
additional power by allowing accesses during refresh. To limit the
power consumption due to \acts, we further constrain the activity rate
by increasing both \tfaw and \trrd, as shown below. This results in
fewer \act commands issued during refresh.




{
\small
\vspace{-0.15in}
\begin{align}
\label{eq:1}
 &PowerOverhead_{FAW} = \frac{4*I_{ACT} + I_{REF}}{4*I_{ACT}} \\
\label{eq:2}
 &t_{FAW\_SARP} = t_{FAW} * PowerOverhead_{FAW} \\
\label{eq:3}
 &t_{RRD\_SARP} = t_{RRD} * PowerOverhead_{FAW}
\end{align}
\vspace{-0.2in}
}


{\small $I_{ACT}$} and {\small $I_{REF}$} represent the current values
of an \act and a refresh, respectively, based on the Micron
Power Calculator~\cite{micron-tr}. We calculate the power overhead of
parallelizing a refresh over a {\em four activate window} using
\eqref{eq:1}. Then we apply this power overhead to both \tfaw
\eqref{eq:2} and \trrd \eqref{eq:3}, which are enforced during refresh
operations. Based on the $IDD$ values in the Micron 8Gb
DRAM~\cite{micronDDR3_8Gb} data sheet, \is increases \tfaw and \trrd
by 2.1x during all-bank refresh operations. Each per-bank refresh consumes 8x
lower current than an all-bank refresh, thus increasing \tfaw and \trrd by only
13.8\%.


\subsubsection{Die Area Overhead} In our evaluations, we use 8 subarrays per
bank and 8 banks per DRAM chip. Based on this configuration, we calculate the
area overhead of \is using parameters from a Rambus DRAM model at 55$nm$
technology~\cite{rambus_powermodel}, the best publicly available model that we
know of, and find it to be 0.71\% in a 2Gb DDR3 DRAM chip with a die area of
73.5$mm^2$. The power overhead of the additional components is negligible
compared to the entire DRAM chip.

\ignore{
\subsection{Summary} We  proposed two mechanisms to hide
refresh latency by parallelizing refresh operations with memory
accesses at different granularities within the DRAM architecture. Our
first mechanism, \darp, exploits the refresh scheduling flexibility in
JEDEC standards to enable the memory controller to make more
intelligent decisions on when and which specific bank to refresh. The
second mechanism, \is, pushes the boundaries of today's DRAMs to
enable a DRAM to serve memory requests in idle subarrays while another
subarray is under refresh. As a result, it requires modifications to
the existing DRAM microarchitecture to achieve refresh-access
parallelization without impacting the array structures.  \is is
complementary to \darp and the combination of both, called \combo,
offers additive performance benefits with the implementation cost
incurred from both mechanisms.}

\section{Methodology}
\label{methodology}

\figputGTS{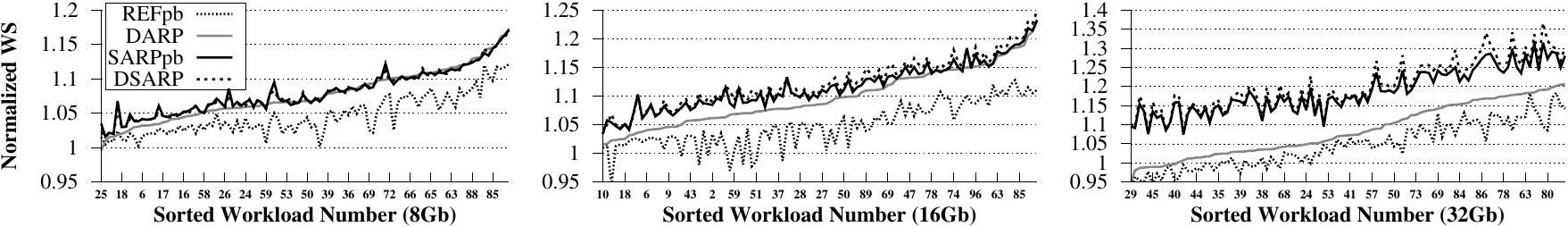}{1.0}{Multi-core system performance improvement over
\caprefab across 100 workloads.}

To evaluate our mechanisms, we use an in-house cycle-level x86
multi-core simulator with a front end driven by
Pin~\cite{luk-pldi2005} and an in-house cycle-accurate DRAM timing
model validated against DRAMSim2~\cite{rosenfeld-cal2011}. Unless
stated otherwise, our system configuration is as shown in
Table~\ref{table:sys-config}.


%
%
%
%

\begin{table}[h]
\begin{footnotesize}
  \centering
    \setlength{\tabcolsep}{.55em}
    \begin{tabular}{ll}
        \toprule
\multirow{2}{*}{Processor} &
8 cores, 4GHz, 3-wide issue, 8 MSHRs/core,\\
& 128-entry instruction window
\\

        \cmidrule(rl){1-2}
\multirow{2}{*}{\begin{minipage}{0.5in}Last-level Cache\end{minipage}} &
64B cache-line, 16-way associative,\\
& 512KB private cache-slice per core \\

        \cmidrule(rl){1-2}
\multirow{3}{*}{\begin{minipage}{0.5in}Memory Controller\end{minipage}} &
64/64-entry read/write request queue, FR-FCFS~\cite{rixner-isca2000}, \\
& writes are scheduled in batches~\cite{chatterjee-hpca2012, lee-tech2010, stuecheli-isca2010}
with \\
& low watermark = 32, closed-row policy~\cite{chatterjee-hpca2012, kim-micro2010, rixner-isca2000} \\

        \cmidrule(rl){1-2}
        \multirow{2}{*}{DRAM} & DDR3-1333~\cite{micronDDR3_8Gb}, 2 channels, 2 ranks per
        channel,\\
            & 8 banks/rank, 8 subarrays/bank, 64K rows/bank, 8KB rows \\

        \cmidrule(rl){1-2}

\multirow{2}{*}{\begin{minipage}{0.5in}Refresh Settings\end{minipage}}
        & $tRFC_{ab}$ = 350/530/890ns for 8/16/32Gb
        DRAM chips,\\
        & $tREFI_{ab}$ = 3.9\micro s, $tRFC_{ab}$-to-$tRFC_{pb}$ ratio = 2.3  \\

        \bottomrule
    \end{tabular}
  \caption{Evaluated system configuration.}
  \label{table:sys-config}%
\end{footnotesize}
\end{table}



In addition to 8Gb DRAM, we also evaluate systems using 16Gb and 32Gb
near-future DRAM chips~\cite{itrs-dram}. Because commodity DDR DRAM does not
have support for \refpb, we estimate the \trfcpb values for DDR3 based on the
ratio of \trfc to \trfcpb in LPDDR2~\cite{micronLPDDR2_2Gb} as described in
Section~\ref{sec:refresh-challenge}. We evaluate our systems with 32ms retention
time, which is a typical setting for a server environment and LPDDR DRAM, as
also evaluated in previous work~\cite{nair-hpca2013,stuecheli-micro2010}.

We use benchmarks from \emph{SPEC CPU2006~\cite{spec2006}, STREAM~\cite{stream},
TPC~\cite{tpc}}, and a microbenchmark with random-access behavior similar to
HPCC RandomAccess~\cite{randombench}. We classify each benchmark as either
memory intensive (MPKI $\ge$ 10) or memory non-intensive (MPKI $<$ 10). We then
form five intensity categories based on the fraction of memory intensive
benchmarks within a workload: 0\%, 25\%, 50\%, 75\%, and 100\%. Each category
contains 20 randomly mixed workloads, totaling to 100 workloads for our main
evaluations. For sensitivity studies in Sections~\ref{sec:sense_core},
\ref{sec:sense_tfaw}, \ref{sec:sense_sa}, and \ref{sec:sense_ref}, we run 16
randomly selected memory-intensive workloads using 32Gb DRAM to observe the
performance trend.

We measure system performance with the commonly-used \emph{weighted
  speedup (WS)}~\cite{eyerman-ieeemicro2008,snavely-asplos2000}
metric. To report the DRAM system power, we use the methodology from
the \emph{Micron power calculator}~\cite{micron-tr}. The DRAM device
parameters are obtained from~\cite{micronDDR3_8Gb}. Every workload
runs for 256 million cycles to ensure the same number of refreshes. We
report DRAM system power as \emph{energy per memory access serviced}
to fairly compare across different workloads.

\section{Evaluation}
\label{sec:evaluation}

In this section, we evaluate the performance of the following
mechanisms: 1) the \emph{all-bank} refresh scheme (\refab), 2) the
\emph{per-bank} refresh scheme (\refpb), 3) elastic
refresh~\cite{stuecheli-micro2010}, 4) our first mechanism, \ib, 5)
our second mechanism, \is, that is applied to either \refab (\isab) or
\refpb (\ispb), 6) the combination of \ib and \ispb, called \combo,
and 7) an ideal scheme that eliminates refresh. Elastic
refresh~\cite{stuecheli-micro2010} takes advantage of the refresh
scheduling flexibility in the DDR standard: it postpones a refresh if
the refresh is predicted to interfere with a demand request, based on
a prediction of how long a rank will be idle, i.e., without any demand
request (see Section~\ref{sec:related} for more detail).






\subsection{Multi-Core Results}
\label{sec:multi-core-results}


\figref{scurves} plots the system performance improvement of \refpb, \ib, \ispb,
and \combo over the all-bank refresh baseline (\refab) using various densities
across 100 workloads (sorted based on the performance improvement due to \ib).
The x-axis shows the sorted workload numbers as categorized into five
memory-intensive groups with 0 to 19 starting in the least memory-intensive
group and 80 to 99 in the most memory-intensive one.
Table~\ref{table:scurve_summary} shows the maximum and geometric mean of system
performance improvement due to our mechanisms over \refpb and \refab for
different DRAM densities. We draw five key conclusions from these results.


\ignore{
First, \ib provides system performance gains over both \refpb and
\refab schemes in different DRAM densities with up to 11.0\% and
23.1\% improvement in 16Gb DRAM, respectively. The reason is that \ib
hides refresh latency with writes and issues refresh commands in
out-of-order fashion to reduce refresh interference on reads. Second,
\ispb provides significant system performance improvement over \ib and
refresh baselines for all the evaluated DRAM density as \ispb enables
accesses to idle subarrays in the refreshing banks. \ispb improves
system performance by up to 21.5\% and 28.0\% over \refpb and \refab
in 32Gb DRAM, respectively.  Third, as DRAM density increases, the
performance benefit of \ispb over \ib gets larger. This is because
longer refresh latency becomes more difficult to hide behind writes or
idle banks for \ib. This is also the reason why the performance
improvement due to \ib drops slightly at 32Gb compared to 16Gb.  On
the other hand, \ispb is able to allow a long-refreshing bank to serve
some memory requests in its subarrays.}

First, \ib provides system performance gains over both \refpb and
\refab schemes: 2.8\%/4.9\%/3.8\% and 7.4\%/9.8\%/8.3\% on average in
8/16/32Gb DRAMs, respectively. The reason is that \ib hides refresh
latency with writes and issues refresh commands in out-of-order
fashion to reduce refresh interference on reads. Second, \ispb
provides significant system performance improvement over \ib and
refresh baselines for all the evaluated DRAM densities as \ispb
enables accesses to idle subarrays in the refreshing banks.  \ispb's
average system performance improvement over \refpb and \refab is
3.3\%/6.7\%/13.7\% and 7.9\%/11.7\%/18.6\% in 8/16/32Gb DRAMs,
respectively.  Third, as density increases, the performance
benefit of \ispb over \ib gets larger. This is because the longer
refresh latency becomes more difficult to hide behind writes or idle
banks for \ib. This is also the reason why the performance improvement
due to \ib drops slightly at 32Gb compared to 16Gb.  On the other
hand, \ispb is able to allow a long-refreshing bank to serve some
memory requests in its subarrays.



Fourth, combining both \ispb and \darp (\combo) provides additive system
performance improvement by allowing even more parallelization of
refreshes and memory accesses. As DRAM density (refresh latency)
increases, the benefit becomes more apparent, resulting in improvement
up to 27.0\% and 36.6\% over \refpb and \refab in 32Gb DRAM,
respectively.

Fifth, \refpb performs worse than \refab for some workloads (the
curves of \refpb dropping below one) and the problem is exacerbated
with longer refresh latency. Because \refpb commands cannot overlap
with each other~\cite{jedec-lpddr3}, their latencies are
serialized. In contrast, \refab operates on every bank in parallel, which is
triggered by a single command that partially overlaps refreshes across
different banks~\cite{mukundan-isca2013}. Therefore, in a pathological
case, the \refpb latency for refreshing every bank (eight in most
DRAMs) in a rank is {\small$8 \times \trfcpb = 8 \times
  \frac{\trfc}{2.3} \approx 3.5 \times \trfc$}, whereas all-bank
refresh takes \trfc (see Section~\ref{sec:refresh-challenge}). If a
workload cannot
effectively utilize multiple banks during a per-bank refresh operation, \refpb
may potentially degrade system performance compared to \refab.

\begin{table}[h]
\small
\centering
\renewcommand{\arraystretch}{0.7}
    \begin{tabular}{llrrrr}
      \toprule
      \multirow{2}{*}{\textbf{Density}} &
      \multirow{2}{*}{\textbf{Mechanism}} &
      \multicolumn{2}{c}{\textbf{Max (\%)}} &
      \multicolumn{2}{c}{\textbf{Gmean (\%)}} \\
      & & \refpb & \refab & \refpb & \refab \\
      \midrule

      \multirow{3}{*}{8Gb} & {\ib} & 6.5 & 17.1 & 2.8 & 7.4\\
      &{\ispb} & 7.4 & 17.3 & 3.3 & 7.9\\
      &{\combo} & 7.1 & 16.7 & 3.3 & 7.9\\
      \midrule

      \multirow{3}{*}{16Gb} & {\ib} & 11.0 & 23.1& 4.9& 9.8\\
      &{\ispb} & 11.0 & 23.3 & 6.7 & 11.7\\
      &{\combo} & 14.5 & 24.8 & 7.2 & 12.3\\
      \midrule

      \multirow{3}{*}{32Gb} & {\ib} & 10.7 & 20.5 & 3.8 & 8.3\\
      &{\ispb} & 21.5 & 28.0 & 13.7 & 18.6\\
      &{\combo} & 27.0 & 36.6 & 15.2 & 20.2\\

      \bottomrule
    \end{tabular}
\caption{Maximum and average WS improvement due to our mechanisms
over \caprefpb and \caprefab.}
\label{table:scurve_summary}%
\end{table}

\subsubsection{All Mechanisms' Results}
Figure~\ref{fig:multi-core-perf-density-sweep} shows the average
performance improvement due to all the evaluated refresh mechanisms
over \refab.  The weighted speedup value for \refab is 5.5/5.3/4.8
using 8/16/32Gb DRAM density.  We draw three major conclusions. First,
using \is on all-bank refresh (\isab) also significantly improves
system performance. This is because \is allows a rank to continue
serving memory requests while it is refreshing. Second, elastic
refresh does not substantially improve performance, with an average of
1.8\% over all-bank refresh.  This is because elastic refresh does not
attempt to pull in refresh opportunistically, nor does it try to
overlap refresh latency with other memory accesses. The observation is
consistent with prior work~\cite{nair-hpca2013}. Third, \combo
captures most of the benefit of the ideal baseline ("No REF"),
performing within 0.9\%, 1.2\%, and 3.7\% of the ideal for 8, 16, and
32Gb DRAM, respectively.

\figputGHS{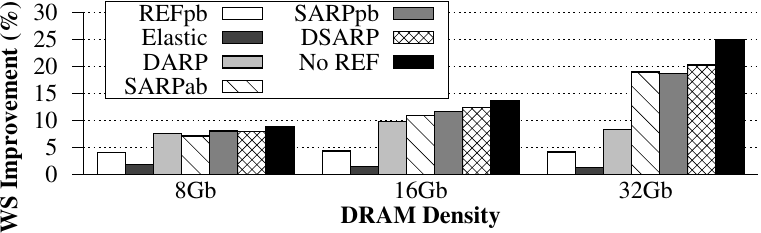}{1.0}{Average system performance
improvement over \caprefab.}

\subsubsection{Performance Breakdown of \ib} To understand the observed
performance gain in more detail, we evaluate the performance of \darp's two
components separately. \emph{\Ooolong} improves performance by 3.2\%/3.9\%/3.0\%
on average and up to 16.8\%/21.3\%/20.2\% compared to \refab in 8/16/32Gb DRAMs.
Adding \emph{\warplong} to \emph{\ooolong} (\ib) provides additional performance
gains of 4.3\%/5.8\%/5.2\% on average by hiding refresh latency with write
accesses.





\subsubsection{Energy}

Our techniques reduce energy per memory access compared to existing
policies, as shown in \figref{energy_reduction}. The main reason is
that the performance improvement reduces average static energy for
each memory access. Note that these results conservatively assume the
same power parameters for 8, 16, and 32 Gb chips, so the savings in
energy would likely be more significant if realistic power parameters
are used for the more power-hungry 16 and 32 Gb nodes.

\figputGHS{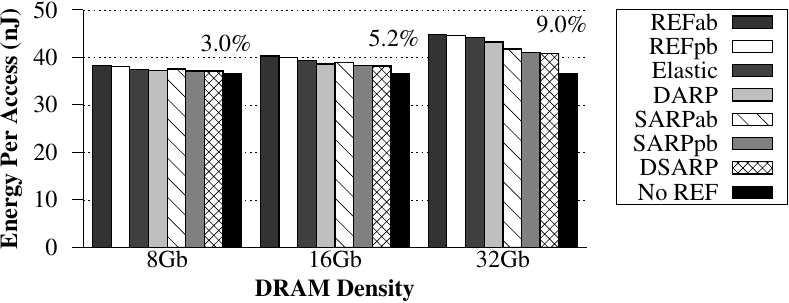}{1.0}{Energy consumption. Value on top indicates
percentage reduction of \combo compared to
\caprefab.}

\subsubsection{Effect of Memory Intensity}
Figure~\ref{fig:mpki_ws} shows the performance improvement of \combo
compared to \refab and \refpb on workloads categorized by memory
intensity (\% of memory-intensive benchmarks in a workload),
respectively. We observe that \combo outperforms \refab and \refpb
consistently. Although the performance improvement of \combo over
\refab increases with higher memory intensity, the gain over \refpb
begins to plateau when the memory intensity grows beyond 25\%. This is
because \refpb's benefit over \refab also increases with memory
intensity as \refpb enables more accesses to be be parallelized with
refreshes. Nonetheless, our mechanism provides the highest system
performance compared to prior refresh policies.

\figputGHS{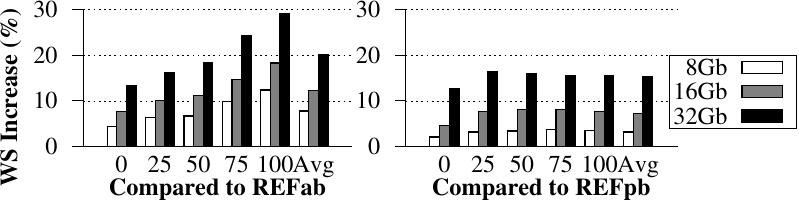}{1.0}{WS improvement of {\small \combo}
over {\small \caprefab} and {\small \caprefpb} as memory intensity and DRAM density vary.}


\subsubsection{Effect of Core Count}
\label{sec:sense_core}

Table~\ref{table:core_sweep} shows the weighted speedup, harmonic speedup,
fairness, and energy-per-access improvement due to \combo compared to \refab for
systems with 2, 4, and 8 cores. For all three systems, \combo consistently
outperforms the baseline without unfairly penalizing any specific application.
We conclude that \combo is an effective mechanism to improve performance,
fairness and energy of multi-core systems employing high-density DRAM.

\begin{table}[h]
\small
\centering
    \setlength{\tabcolsep}{.35em}
    \begin{tabular}{lrrr}
      \toprule
      \textbf{Number of Cores} & 2 & 4 & 8 \\
      \midrule
      \textbf{Weighted Speedup Improvement (\%)} & 16.0 & 20.0 & 27.2 \\
      \textbf{Harmonic Speedup Improvement~\cite{luo-ispass2001} (\%)} & 16.1 & 20.7 & 27.9 \\
      \textbf{Maximum Slowdown Reduction~\cite{das-micro2009, kim-hpca2010,
      kim-micro2010} (\%)} & 14.9 & 19.4 & 24.1 \\
      \textbf{Energy-Per-Access Reduction (\%)} & 10.2 & 8.1 & 8.5 \\
      \bottomrule
    \end{tabular}
\caption{Effect of \combo on multi-core system metrics.}
\label{table:core_sweep}
\end{table}


\subsection{Effect of \tfaw}
\label{sec:sense_tfaw}
Table~\ref{table:faw_sweep} shows the performance improvement of \ispb
over \refpb when we vary \tfaw in DRAM cycles (20 cycles for the
baseline as specified by the data sheet) and when \trrd scales
proportionally with \tfaw.\footnote{We evaluate only \ispb because it
  is sensitive to \tfaw and \trrd as it extends these parameters
  during parallelization of refreshes and accesses to compensate for
  the power overhead.} As \tfaw reduces, the performance benefit of
\ispb increases over \refpb. This is because reduced \tfaw enables
more accesses/refreshes to happen in parallel, which our mechanism
takes advantage of.

\begin{table}[h]
\small
\centering
    \setlength{\tabcolsep}{.33em}
    \begin{tabular}{lrrrrrr}
      \toprule
      \textbf{\tfaw/\trrd} (DRAM cycles) & 5/1 & 10/2 & 15/3 & \textbf{20/4} &
      25/5 & 30/6\\
      \midrule
      \textbf{WS Improvement (\%)} & 14.0 & 13.9 & 13.5 & \textbf{12.4} & 11.9 & 10.3\\
      \bottomrule
    \end{tabular}
\caption{Performance improvement due to \ispb over \caprefpb with various
\tfaw and \trrd values.}
\label{table:faw_sweep}
\end{table}

\vspace{-0.08in}

\subsection{Effect of Subarrays-Per-Bank}
\label{sec:sense_sa}
Table~\ref{table:sa_sweep} shows that the average performance
gain of \ispb over \refpb increases as the number of subarrays
increases in 32Gb DRAM. This is because with more subarrays, the
probability of memory requests to a refreshing subarray reduces.

\begin{table}[h]
\small
\centering
    \setlength{\tabcolsep}{.33em}
    \begin{tabular}{lrrrrrrr}
      \toprule
      \textbf{Subarrays-per-bank} & 1 & 2 & 4 & \textbf{8} & 16 & 32 & 64\\
      \midrule
      \textbf{WS Improvement (\%)} & 0 & 3.8 & 8.5 & \textbf{12.4} & 14.9 & 16.2
      & 16.9\\
      \bottomrule
    \end{tabular}
\caption{Effect of number of subarrays per bank.}
\label{table:sa_sweep}
\end{table}

\vspace{-0.08in}

\subsection{Effect of Refresh Interval}
\label{sec:sense_ref}

For our studies so far, we use 32ms retention time (i.e., \trefi =
3.9$\mu$s) that represents a typical setting for a server environment
and LPDDR DRAM~\cite{jedec-lpddr3}. Table~\ref{table:64ms_summary}
shows the performance improvement of \combo over two baseline refresh
schemes using retention time of \emph{64ms} (i.e., \trefipb =
7.8$\mu$s). \combo consistently provides performance gains over both
refresh schemes. The maximum performance improvement over \refpb is
higher than that over \refab at 32Gb because \refpb actually degrades
performance compared to \refab for some workloads, as discussed in the
32ms results (Section~\ref{sec:multi-core-results}).

\begin{table}[h]
\small
\centering
\renewcommand{\arraystretch}{0.65}
    \begin{tabular}{lrrrr}
      \toprule
      \multirow{2}{*}{\textbf{Density}} &
      \multicolumn{2}{c}{\textbf{Max (\%)}} &
      \multicolumn{2}{c}{\textbf{Gmean (\%)}} \\
      & \refpb & \refab & \refpb & \refab \\
      \midrule

      {8Gb} &  2.5   & 5.8 & 1.0& 3.3 \\
      {16Gb}&  4.6   & 8.6 & 2.6& 5.3 \\
      {32Gb} & 18.2  & 13.6& 8.0& 9.1 \\

      \bottomrule
    \end{tabular}
\caption{Maximum and average WS improvement due to \combo.}
\label{table:64ms_summary}%
\end{table}

\subsection{DDR4 Fine Granularity Refresh}
\label{sec:ddr4}

DDR4 DRAM supports a new refresh mode called \emph{fine granularity
  refresh (FGR)} in an attempt to mitigate the increasing refresh
latency (\trfc)~\cite{jedec-ddr4}. \fgr trades off shorter \trfc with
faster refresh rate ($\sfrac{1}{\trefi}$) that increases by either 2x
or 4x. \figref{ddr4_ref} shows the effect of \fgr in comparison to
\refab, \emph{adaptive refresh policy (AR)}~\cite{mukundan-isca2013},
and \combo. 2x and 4x \fgr actually reduce average system performance
by 3.9\%/4.0\%/4.3\% and 8.1\%/13.7\%/15.1\% compared to \refab with
8/16/32Gb densities, respectively. As the refresh rate increases by
2x/4x (higher refresh penalty), \trfc does not scale down with the
same constant factors. Instead, \trfc reduces by 1.35x/1.63x with
2x/4x higher rate~\cite{jedec-ddr4}, thus increasing the worst-case
refresh latency by 1.48x/2.45x. This performance degradation due to
\fgr has also been observed in Mukundan et
al.~\cite{mukundan-isca2013}.  AR~\cite{mukundan-isca2013} dynamically
switches between 1x (i.e., \refab) and 4x refresh modes to mitigate
the downsides of \fgr.  AR performs slightly worse than \refab (within
1\%) for all densities. Because using 4x \fgr greatly degrades
performance, AR can only mitigate the large loss from the 4x mode and
cannot improve performance over \refab. On the other hand, \combo is a
more effective mechanism to tolerate the long refresh latency than
both \fgr and AR as it overlaps refresh latency with access latency
without increasing the refresh rate.


\figputGHS{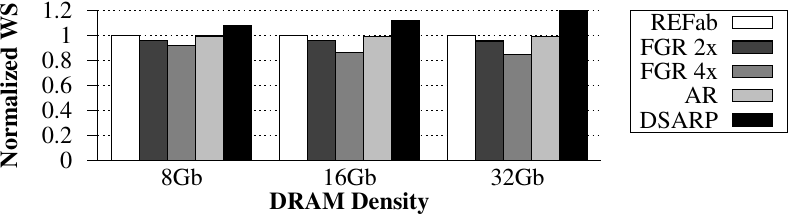}{1.0}{Performance comparisons to FGR and AR~\cite{mukundan-isca2013}.}

\section{Related Work}
\label{sec:related}

To our knowledge, this is the first work to comprehensively study the
effect of per-bank refresh and propose 1) a refresh scheduling policy
built on top of per-bank refresh and 2) a mechanism that achieves
parallelization of refresh and memory accesses {\em within} a
refreshing bank. We discuss prior works that mitigate the negative
effects of DRAM refresh and compare them to our mechanisms.

\textbf{Refresh Scheduling.} Stuecheli et al.~\cite{stuecheli-micro2010}
propose elastic refresh, which we discussed and evaluated in
Section~\ref{sec:evaluation}. Elastic refresh postpones refreshes by a time
delay that varies based on the number of postponed refreshes and the predicted
rank idle time to avoid interfering with demand requests. Elastic refresh has
two shortcomings. First, it becomes less effective when the average rank idle
period is shorter than \trfc as the refresh latency cannot be fully hidden in
that period. This occurs especially with 1) more memory-intensive workloads
that inherently have less idleness and 2) higher density DRAM chips that have
higher \trfc. Second, elastic refresh incurs more refresh latency when it {\em
incorrectly} predicts a time period as idle when it actually has pending
requests. In contrast, our mechanisms parallelize refresh operations with
accesses even if there is no idle period and therefore outperform elastic
refresh.


Ishii et al.~\cite{ishii-msc2012} propose a write scheduling policy
that prioritizes write draining over read requests in a rank while
another rank is refreshing (even if the write queue has not reached
the threshold to trigger write mode). This technique is {\em only}
applicable in multi-ranked memory systems.  Our mechanisms are {\em
  also} applicable to single-ranked memory systems by enabling
parallelization of refreshes and accesses at the bank and subarray
levels, and they can be combined with Ishii et
al.~\cite{ishii-msc2012}.




Mukundan et al.~\cite{mukundan-isca2013} propose scheduling techniques (in
addition to adaptive refresh discussed in Section~\ref{sec:ddr4}) to address the
problem of {\em command queue seizure}, whereby a command queue gets filled up
with commands to a refreshing rank, blocking commands to {\em another}
non-refreshing rank.  In our work, we use a different memory controller design
that does not have command queues, similarly to prior
work~\cite{herrero-tc2013}. Our controller generates a command for a scheduled
request {\em right before} the request is sent to DRAM instead of pre-generating
the commands and queueing them up. Thus, our baseline design does not suffer
from the problem of command queue seizure.


\textbf{Subarray-Level Parallelism (SALP).} Kim et
al.~\cite{kim-isca2012} propose SALP to reduce bank serialization
latency by enabling {\em multiple accesses} to different subarrays
within a bank to proceed in a pipelined manner. In contrast to SALP,
our mechanism (\is) parallelizes {\em refreshes and accesses} to
different subarrays within the same bank. Therefore, \is exploits the
existence of subarrays for a different purpose and in a different way
from SALP. As Section~\ref{sec:sarp_implementation} describes, we
reduce the sharing of the peripheral circuits for refreshes and
accesses, not for arbitrary accesses. As such, our implementation is
not only different, but also less intrusive than SALP: \is does not
require new DRAM commands and timing constraints.

\textbf{Refresh Pausing.} Nair et al.~\cite{nair-hpca2013} propose
pausing a refresh operation to serve pending memory requests. To make
pausing a refresh possible, the authors assume that DRAM refreshes
multiple rows \emph{sequentially}. Thus, there is a short recovery
time, called a "refresh pausing point" (RPP), after refreshing each
row so that the memory controller can signal the DRAM to stop
refreshing subsequent rows. However, DRAM manufacturers currently
design their chips to refresh multiple rows in \emph{parallel} (or, in
a {\em staggered/pipelined} way, as explained
in~\cite{mukundan-isca2013}).

\textbf{eDRAM Concurrent Refresh.} Kirihata et al.~\cite{kirihata-jssc2005}
propose a mechanism to enable a bank to refresh independently while another bank
is being accessed in embedded DRAM (eDRAM). Our work differs
from~\cite{kirihata-jssc2005} in two major ways. First, unlike
SARP,~\cite{kirihata-jssc2005} parallelizes refreshes only across banks, not
{\em within} each bank.  Second, there are significant differences between DRAM
and eDRAM architectures, which make it non-trivial to
apply~\cite{kirihata-jssc2005}'s mechanism directly to DRAM. In particular,
eDRAMs have no standardized timing/power integrity constraints and access
protocol, making it simpler for each bank to independently manage its refresh
schedule. In contrast, refreshes in DRAM need to be managed by the memory
controller to ensure that parallelizing refreshes with accesses does not violate
other constraints.


\textbf{Retention-Aware Refresh.} Prior works
(e.g.,~\cite{liu-isca2012, venkatesan-hpca2006}) propose mechanisms to
reduce unnecessary refresh operations by taking advantage of the fact
that different DRAM cells have widely different retention
times~\cite{liu-isca2013}. These works assume that the retention time
of DRAM cells can be \emph{accurately} profiled and they depend on
having this accurate profile in order to guarantee data
integrity~\cite{liu-isca2013}. However, as shown
in~\cite{liu-isca2013}, accurately determining the retention time
profile of DRAM is an unsolved research problem due to the Variable
Retention Time and Data Pattern Dependence phenomena, which can cause
the retention time of a cell to fluctuate over time.  As such,
retention-aware refresh techniques need to overcome the profiling
challenges to be viable.



\section{Conclusion}

We introduced two new complementary techniques, \ib (\darplong) and
\is (\intersub), to mitigate the DRAM refresh penalty by enhancing
\emph{refresh-access parallelization} at the bank and subarray levels,
respectively.  \darp 1) issues per-bank refreshes to idle banks in an
out-of-order manner instead of issuing refreshes in a strict
round-robin order, 2) proactively schedules per-bank refreshes during
intervals when a batch of writes are draining to DRAM.  \is enables a
bank to serve requests from idle subarrays in parallel with other
subarrays that are being refreshed. Our extensive evaluations on a
wide variety of systems and workloads show that these mechanisms
significantly improve system performance and outperform
state-of-the-art refresh policies, approaching the performance of
ideally eliminating all refreshes. We conclude that \ib and \is are
effective in hiding the refresh latency penalty in modern and
near-future DRAM systems, and that their benefits increase as DRAM
density increases.


}

\vspace{-0.06in}
\section*{Acknowledgments}
\vspace{-0.07in}
{We thank the anonymous reviewers and Jamie Liu for
helpful feedback and the members of the SAFARI research group for feedback and
the stimulating environment they provide. We acknowledge the support of IBM,
Intel, and Samsung. This research was supported in part by the Intel Science and
Technology Center on Cloud Computing,
SRC (Semiconductor Research Corporation), and an NSF CAREER Award 0953246.}

\vspace{-0.06in}
{
\small
\bstctlcite{bstctl:etal, bstctl:nodash, bstctl:simpurl}
\bibliographystyle{IEEEtranS}
\bibliography{paper}
}

\newpage
\section*{Erratum}

Our HPCA 2014 paper contains an error on the description
of the \textit{Dynamic Access Refresh Parallelization} mechanism in Section 4.2,
on the left column of page 5.

\textbf{Error}:
``Therefore, a \refpb command can be postponed if the bank's ref\_credit
stays above -8. Otherwise the memory controller is required to send a
REFpb command to comply with the standard.''

\textbf{Correction}:
``Therefore, a \refpb command can be postponed if the bank's ref\_credit
stays between values of 0 and 8 ($0\le ref\_credit \le8$). Otherwise the
memory controller is required to send a \refpb command when more than eight
  \refpb commands have been postponed (i.e., $8 < ref\_credit$) to comply with
the standard. Each \refpb resets the bank's ref\_credit value bank to 0.''

\textbf{Reason}: The original paper states that a per-bank refresh command
(\refpb) can be postponed when the bank's ref\_credit stays between -8 and 8.
Setting a lower bound at -8 can potentially cause a bank to operate without any
refreshes for more than 9 refresh intervals (i.e., 9 \trefipb). This would
violate the JEDEC standard requirement of allowing a bank to
skip a maximum of 9 \trefipb without refreshes in order to guarantee data
integrity in DRAM. To prevent such violations to occur, we set the range of each
bank's ref\_credit from 0 to 8. This guarantees that each bank skips up to only
8 refresh commands (\refpb) at any time. Once a bank is refreshed, the memory
controller resets the bank's ref\_credit back to 0.

\end{document}